\documentclass[a4paper,11pt]{article}
\pdfoutput=1

\usepackage{jcappub}

\usepackage{bm}
\usepackage{enumerate}
\usepackage{cleveref}
\crefname{figure}{figure}{figures}
\crefname{table}{table}{tables}
\def\equationautorefname~#1\null{%
  eq.~(#1)\null
}
\def\figureautorefname~#1\null{%
  figure~#1\null
}
\def\tableautorefname~#1\null{%
  table~#1\null
}
\usepackage[dvipsnames]{xcolor}

\title{\boldmath
   Post-inflationary axion isocurvature perturbations facing CMB and large-scale structure
}

\author[a,b,c]{M. Feix,}
\author[d]{S. Hagstotz,}
\author[c,e]{A. Pargner,}
\author[f,g]{R. Reischke,}
\author[b,c]{\newline B.M. Sch\"afer,}
\author[c,e]{T. Schwetz}

\affiliation[a]{Zentrum f{\"u}r Astronomie der Universit{\"a}t Heidelberg, Institut f{\"u}r Theoretische Astrophysik, \\ Philosophenweg 12, 69120 Heidelberg, Germany}
\affiliation[b]{Zentrum f{\"u}r Astronomie der Universit{\"a}t Heidelberg, Astronomisches Rechen-Institut,\\ Philosophenweg 12, 69120 Heidelberg, Germany}
\affiliation[c]{HEiKA -- Heidelberg Karlsruhe Research Partnership, Heidelberg University,\\ Karlsruhe Institute of Technology (KIT), Germany}
\affiliation[d]{The Oskar Klein Centre for Cosmoparticle Physics, Department of Physics, Stockholm University, Roslagstullsbacken 21A, SE-106 91 Stockholm, Sweden}
\affiliation[e]{Institut f\"ur Astroteilchenphysik, Karlsruher Institut f\"ur Technologie (KIT),\\ Hermann-von-Helmholtz-Platz 1, 76344 Eggenstein-Leopoldshafen, Germany}
\affiliation[f]{Department of Physics, Israel Institute of Technology -- Technion,\\ 3200003 Haifa, Israel}
\affiliation[g]{Department of Natural Sciences, The Open University of Israel, 1 University Road, P.O. Box 808, Ra'anana 4353701, Israel}

\emailAdd{steffen.hagstotz@fysik.su.se}
\emailAdd{r.reischke@campus.technion.ac.il}

\abstract{ Dark matter comprised of axion-like particles (ALPs)
  generated by the realignment mechanism in the post-inflationary
  scenario leads to primordial isocurvature fluctuations. The power
  spectrum of these fluctuations is flat for small wave
  numbers, extending to scales accessible with cosmological
  surveys. We use the latest measurements of Cosmic Microwave
  Background (CMB) primary anisotropies (temperature, polarization)
  together with CMB lensing, Baryonic Acoustic Oscillations (BAO) and
  Sunyaev Zel'dovich (SZ) cluster counts to measure the amplitude and
  tilt of the isocurvature component.
  We find preference for a white-noise isocurvature
  component in the CMB primary anisotropies; this conclusion is,
  however, weakened by current large-scale structure (LSS)
  data. Interpreting the result as a conservative upper limit on the isocurvature
  component, the combined bound on the ALP mass from all probes is
  $m_{a} \gtrsim 10^{-19}$~eV, with some dependence on how $m_{a}$
  evolves with temperature. The expected sensitivity of cosmic shear
  and galaxy clustering from future LSS experiments and CMB lensing
  suggests improved bounds of $m_{a} \gtrsim 10^{-18}$--$10^{-13}$~eV,
  depending on scale cuts used to avoid non-linearities and the ALP
  mass-temperature dependence.}

\keywords{axions, cosmological parameters from LSS, dark matter theory}

\begin{document}
\maketitle
\flushbottom

\section{Introduction}
\label{sec:intro}

If interpreted within the framework of general relativity, astrophysical and cosmological data are supporting the existence of dark matter (DM)
\citep{Bertone2005, Bertone2018}. Together with the evidence for an accelerated expansion of the Universe \citep{riess_observational_1998, perlmutter_measurements_1999,
riess_new_2007}, this has led to the widely accepted cold DM cosmology with a cosmological constant, the so-called $\Lambda$CDM model \citep[e.g.,][]{bartelmann_dark_2010}.
Still, DM remains merely a postulate and constitutes one of the biggest puzzles in fundamental physics.

Over the years, a variety of DM candidates have been proposed \citep[see, e.g.,][]{bauer_yet_2017}. A specific and well-motivated particle physics
candidate for DM is the axion \citep{Weinberg1978, Wilczek1978, Sikivie2008, Marsh2016}. It appears as a pseudo-Nambu-Goldstone boson (PNGB) in the Peccei-Quinn (PQ)
solution to the strong $\mathrm{CP}$ problem \citep{Peccei1977}. The axion emerges from a new global chiral $U(1)$ symmetry that gets spontaneously broken at an energy
scale $f_{a}$ by the vacuum expectation value of a complex scalar field. The scale $f_{a}$ may be assumed very large to satisfy current experimental bounds, giving rise
to weak interactions and tiny axion masses \citep{Dine1981, Zhitnitsky1980, Kim1979, Shifman1980}. Other high-energy extensions to the standard model of particle physics
containing PNGBs and sharing properties similar to those of the axion are referred to as axion-like particles (ALPs) \citep{Arvanitaki2010, Arias2012, Ringwald2012}.

Due to non-thermal production via the vacuum realignment mechanism \citep{Dine1983, Preskill1983, Turner1983, Abbott1983}, ALPs can mimic a cold DM component despite
their small masses. At high temperatures $T\sim f_{a}$, ALPs are basically massless and assume random field values. At much lower temperatures $T_{\rm osc} \ll f_a$,
a potential develops due to non-perturbative effects, the ALP becomes massive, and coherent oscillations around the minimum of the potential behave like collisionless
cold DM on scales relevant to observations of cosmic large-scale structure (LSS). While the QCD axion requires a mass $m_{a}\sim 10^{-5}$~eV to match the observed DM
density \citep[e.g.,][]{Visinelli2009, Kawasaki:2014sqa, Klaer2017,Gorghetto2018,Vaquero2018}, ALP masses can extend to much smaller values \citep[e.g.,][]{Arias2012,
Grin:2019mub, Niemeyer:2019aqm}. The allowed mass range is, for instance, constrained from structure formation arguments \citep[e.g.,][]{Amendola2006, Marsh2010, Marsh2012,
Akrami2018a, hlozek_future_2016, hlozek_using_2018, Haehnelt2017, Kobayashi2017, Bauer:2020zsj}. In particular, a strong bound is obtained from Lyman-$\alpha$ observations
\citep{Haehnelt2017, Kobayashi2017}:
\begin{equation}\label{eq:bounds}
  m_a \gtrsim   10^{-21} \,\text{eV} \,.
\end{equation}
Under certain astrophysical modelling assumptions, the formation of
solitonic cores in DM halos leads to $m_a \gtrsim
10^{-19}$~eV~\citep{Marsh2018}. Further constraints are given by the
spin-down of black holes via superradiant instability
\citep[e.g.,][]{Arvanitaki2015, Stott2018} which, however, do not
apply in the scenario considered here since for $f_a$-values needed to
match the observed DM abundance, ALP self-interactions prevent the
build-up of the axionic cloud around the black hole \cite{Arvanitaki2015}\footnote{Compare
  the $f_a$ values shown in the left panel of Fig.~1 of
  \cite{Feix2019} with the estimate in Eq.~(10) of
  \cite{Arvanitaki2015}.}.

Depending on whether the PQ symmetry is broken before the end of inflation
or thereafter, the vacuum realignment mechanism yields two different scenarios. In the first case, ALPs act like spectator fields and introduce isocurvature fluctuations that
follow the usual scale-invariant spectrum produced during inflation, providing a 
cosmological test of the pre-inflationary scenario \citep{Turner:1990uz, Lyth:1991ub, Beltran2007,Hertzberg2008, Hamann:2009yf, Visinelli:2017imh, Schmitz:2018nhb}.
In the second scenario, however, the ALP field takes different values in causally disconnected regions, leading to large additional isocurvature fluctuations that are
characterized by a blue spectrum (compared to the one of adiabatic modes). This has interesting consequences such as the formation of gravitationally bound objects known
as miniclusters \citep{Hogan1988, Kolb1993, Kolb1994b, Kolb1996, Zurek2007, Hardy2016, Enander2017, Vaquero2018, Buschmann:2019icd, Eggemeier:2019khm}. The power spectrum of
these isocurvature fluctuations was computed for the QCD axion in \citep{Enander2017} and has been generalized to the case of ALPs in \cite{Feix2019}.
Considering the cosmic microwave background (CMB) \citep{planck_collaboration_planck_2016, Planck2016_likelihoods}
and future HI intensity mapping experiments, first constraints on ALP masses in the post-inflationary symmetry breaking scenario were obtained in \cite{Feix2019}. More
recently, similar bounds have been discussed in the context of the reionization history and Lyman-$\alpha$ observations \citep{Irsic2019}.

This work continues our investigations on ALP DM generated from PQ symmetry breaking after inflation and its imprints on various cosmological probes. After presenting current results from the CMB, from baryon acoustic oscillations (BAO) and the abundance of galaxy clusters, we extend the forecasts from \cite{Feix2019} with a study on how future surveys
focusing on weak gravitational lensing either of galaxies \citep{bacon_detection_2000, brown_measurement_2002, heymans_cfhtlens_2013, hildebrandt_inferring_2013} or of the CMB
\citep{lewis_weak_2006, hirata_reconstruction_2003} can constrain isocurvature perturbations. We will also include galaxy clustering and cross-correlations between the various probes.
For the forecasts presented in this paper we will consider a galaxy survey similar to Euclid
\citep{laureijs_euclid_2009} and a CMB stage-IV experiment \citep{abazajian_cmb-s4_2016}.

We structure the paper as follows: in \cref{sec:alp}, we briefly
introduce the axion model and relevant approximations, including our
treatment of scales where non-linear gravitational dynamics starts to
become important.  In \cref{sec:constraints}, we discuss current
constraints on the isocurvature component and the resulting limits on
the ALP mass from latest measurements of the CMB, of BAOs and cluster counts. \Cref{sec:lss}
summarizes the different LSS probes used in our forecast, and expected future
constraints on the isocurvature mode and associated ALP masses are
presented in \cref{sec:lss-sens}. Finally, we conclude in
\cref{sec:conclusion}.

Throughout, we will assume a spatially flat reference cosmology based on \citep{collaboration_planck_2016}, adopting the total matter density
parameter $\Omega_{\rm m} = 0.315$, the baryon density parameter $\Omega_{\rm b} = 0.049$, the amplitude $A_{\rm s} = 2.215\times 10^{-9}$
of the primordial adiabatic spectrum, its spectral index $n_{\rm s} = 0.9603$ (without running, i.e. $\alpha_{\rm s} = 0$), the optical depth
$\tau = 0.089$, the dimensionless Hubble parameter $h = 0.673$, and the sum of neutrino masses $\sum m_{\nu} = 0.06$ eV. Any other parameters
relevant to our analysis will be introduced and specified below.

\section{ALPs from symmetry breaking after inflation}
\label{sec:alp}
To study the evolution of the ALP field in the post-inflationary PQ breaking scenario, we use the semi-analytic method derived for the QCD axion
in \cite{Enander2017} which was generalized to the case of ALPs in \cite{Feix2019}. Details of the calculation are given in these references. In
what follows, we briefly review the procedure, discuss the most important assumptions, and summarize results necessary for the present analysis.

\subsection{ALP field and cosmic evolution}
\label{sec:alp_cosmic}
Assuming that the potential for the ALP field, $\phi_a(x)$, is generated in a similar fashion as for the QCD axion, we may write
\begin{equation}
V(\theta, T) = m^{2}_{a}(T)\left (1 - \cos\theta\right ),
\label{Eq:Potential}
\end{equation}
where we have introduced the dimensionless realignment field $\theta(x)\equiv \phi_a(x)/f_a$ and $m_{a}(T)$ is the temperature-dependent ALP mass\footnote{$V(\theta, T)$ denotes the potential for the dimensionless field $\theta$, which we define to have mass-dimension 2, such that $V(\theta, T)$ has the same dimension as the kinetic term for $\theta$, see Eq.~(2.4) of \cite{Feix2019}.}.
For the latter, we assume the parametric form
\begin{equation}
m_{a}(T) = \mathrm{min}\left\lbrack m_{a},\; m_{a}b
\left (\frac{\Lambda}{T}\right )^{n}\right\rbrack \,,\qquad m_a = \frac{\Lambda^2}{f_a}\,.
\label{Eq:MassParameterization}
\end{equation}
Moving from high temperatures, $T\gg\Lambda$, to lower ones, the ALP mass emerges through a power law, controlled by the parameter $n\in\mathbb{R}_{+}$,
and reaches its zero-temperature value $m_{a}$ at $T_{0} = b^{1/n}\Lambda$. In analogy to the QCD axion, $\Lambda$ plays the role of a topological
susceptibility in a strongly interacting sector. 
The parameter $b$ takes into account that the mass might not reach $m_{a}$ exactly at $T = \Lambda$.
We take as independent parameters $m_a,f_a,b,n$, where later we will fix one of them (namely $f_a$) by requiring that the ALP energy density provides all DM.
To investigate different ALP scenarios and the impact of the temperature dependence, we will consider values $0.1 < b,n < 10$.

The full potential in \cref{Eq:Potential} leads to a complicated non-linear equation of motion for $\theta(x)$. This gives rise to many interesting
effects such as the formation of topological defects and ultra-compact field configurations \citep{Vaquero2018}. To make analytic progress, however,
we can use the harmonic approximation for small field values, i.e. $V(\theta,T)\simeq m_{a}^2(T)\theta^{2}/2$. Although this ignores all non-linear
effects, it turns out that the results of the calculation are very useful and quite accurate for studying large-scale observables \cite{Feix2019, Pargner:2019wxt}.\footnote{We expect that non-linearities in the potential become important at scales comparable to or smaller than the horizon at $T_{\rm osc}$. For the observables of interest to us, however, much larger scales are relevant, i.e. $k\ll K$ where \cref{Eq:Delta} is valid (see below).}

In the harmonic approximation, the evolution equations for the Fourier modes $\theta_{\mathbf{k}}$ of the realignment field are given by
\begin{equation}
\ddot{\theta}_{\mathbf{k}} + 3H(T)\dot{\theta}_{\mathbf{k}} + \left\lbrack\frac{k^{2}}{a^{2}} + m^{2}_a(T)\right\rbrack\theta_{\mathbf{k}} = 0\;,
\end{equation}
where $H(T)=\dot{a}/a$ is the Hubble rate, $a(t)$ is the scale factor of the cosmological background, and dots denote derivatives with respect to cosmic time $t$.
Since the equations for different $\theta_{\mathbf{k}}$ decouple, the system may be solved mode by mode. For high temperatures above the oscillation
temperature $T_{\mathrm{osc}}$, the solution is found numerically and matched to a WKB approximation at $T< T_{\mathrm{osc}}$. The temperature $T_{\mathrm{osc}}$
is defined as $m_a(T_{\mathrm{osc}})=3H(T_{\mathrm{osc}})$, and approximately equals the time when the zero mode starts to oscillate. 

The result is then used to compute the mean relic ALP energy density
$\overline{\rho}_a$ and the initial power spectrum $P(k)$ of density
fluctuations.  In doing so, we assume that the initial $\theta(x)$
produced by the PQ phase transition is a Gaussian random field, and
thus fully characterized by its power spectrum $P_{\theta}(k)$ which
is defined through
$\langle\theta_{\mathbf{k}}\theta^{\ast}_{\mathbf{k}^{\prime}}\rangle
= (2\pi )^{3}\delta_{D}(\mathbf{k}-\mathbf{k}^{\prime})P_\theta(k)$
where $\delta_D$ denotes the 3-dimensional Dirac-delta function.  In the
post-inflationary scenario considered here, the ALP field $\theta$
assumes uncorrelated values in causally disconnected regions, whereas
gradient terms in the field equations tend to homogenize $\theta$
inside the horizon. The Kibble mechanism \cite{Kibble1976} ensures
that the field remains in such a state at any time after PQ breaking
and well before field oscillations become important. Therefore, $\theta(x)$ is
fully correlated on scales smaller than the causal horizon and
uncorrelated otherwise. To capture this behavior, we assume
$P_{\theta}\propto \exp[-k^2/Q^2]$, yielding a white-noise behavior on
large scales with a smooth cutoff at the characteristic scale
$Q=a_iH(T_i)$, where $T_{i}$ is the initial temperature at which we
begin evolving the field equation. As the neglected non-linear effects
are potentially important for the early field evolution
\cite{Hindmarsh1995}, we cannot set $T_{i}$ as high as the PQ breaking
scale, but rather start close to $T_{\mathrm{osc}}$. As default for
our study, we use $T_i=3T_{\mathrm{osc}}$. Further details of this
approach can be found in \cite{Enander2017}.

\subsection{Initial isocurvature spectrum}
\label{sec:alp_iso}
Fluctuations in the ALP energy density are expected to inherit the white-noise character of the initial random field at large scales corresponding to wave numbers $k$ smaller than
a characteristic scale of order $K = a_{\mathrm{osc}}H(T_{\mathrm{osc}})$. Indeed, departing from the field power spectrum $P_\theta(k)$ discussed above, numerical calculations \cite{Enander2017} show that the energy density power sepctrum $P(k)$ is almost constant
for $k\lesssim K$ and that the dimensionless power spectrum $\Delta^{2} = k^{3}P(k)/2\pi^{2}$ can be parametrized as
\begin{equation}
\Delta^{2}(k)=C\left(\frac{k}{K}\right)^{3} \qquad (k\lesssim K)\,,
\label{Eq:Delta}
\end{equation}
where $C$ is a constant that is sensitive to the temperature-dependent
ALP mass. For the parameter range of $b$ and $n$ considered here, we
obtain $0.04\lesssim C \lesssim 0.3$ \cite{Feix2019}. Compared to
numerical simulations that include the full non-linear potential for
the QCD axion \citep{Vaquero2018}, we find a factor five
difference. Below, we take this factor as a systematic uncertainty on
the predicted value of $C$ when analyzing different ALP models.  Let
us stress that while the behaviour of the power spectrum for $k\gtrsim
K$ is complicated and can be assessed only by numerical simulations,
the white-noise shape for $k\lesssim K$ according to \cref{Eq:Delta}
is a robust prediction, confirmed also by numerical simulations
\citep{Vaquero2018,Gorghetto2018}, since it is based only on causality
arguments. For the parameter space in $(m_a, n, b)$ relevant for our analysis it turns
out that $K$ is in the range $10^1$ to $10^3$~Mpc$^{-1}$ (see right
panel of Fig.~1 of \cite{Feix2019}). As we discuss below, for the
observables of interest in this work we are restricting the analysis
to the linear (or mildly non-linear) regime of fluctuations, which
indeed implies that only scales $k\lesssim K$ are relevant, justifying
the use of \cref{Eq:Delta}.

For lower ALP masses, the field starts oscillating at later times, and the characteristic scale of fluctuations $K$ increases. This means that the impact of
white-noise isocurvature fluctuations in the ALP energy density will become relevant on cosmological scales for extremely light ALPs only. Requiring that
the ALP field comprises all DM, we set the relic ALP density $\overline{\rho}_a$ to the observed DM density $\rho_{\mathrm{DM}}$, which fixes the breaking
scale $f_{a}$ as a function of the zero-temperature mass $m_{a}$. Although there remains some dependence on the parameters $b$ and $n$, very light ALPs
generally need a high breaking scale, e.g., $f_a\sim 10^{16}-10^{17}~\mathrm{GeV}$ for $m_a\sim 10^{-22}~\mathrm{eV}$, see Fig.~1 (left) of Ref.~\cite{Feix2019}. In the post-inflationary PQ breaking
scenario, however, $f_{a}$ should either be bound by the energy scale of inflation or there exists a mechanism restoring the PQ symmetry after inflation
has ended. In this work, we take the view of being agnostic to the exact details of inflation, and allow for PQ breaking scales as high as
$f_{a}\sim 10^{17}~\mathrm{GeV}$. For a more elaborate discussion of the conditions in the post-inflationary scenario, we refer to \cite{Feix2019}.

\subsection{Cold DM approximation and gravitational non-linearity}
\label{sec:alp_cdm_approx}
The initial perturbations described in \cref{sec:alp_iso} are set deep within the radiation era where all modes of interest are well outside the
horizon. For sufficiently large $m_{a}$, the ALP field may be approximated in terms of a standard cold DM component with an additional isocurvature
mode specified by the spectrum \cref{Eq:Delta}. 
%

Following \cite{Feix2019}, the amplitude of ALP DM isocurvature perturbations is expressed relative to the adiabatic mode through the entropy-to-curvature
ratio, $f_{\rm iso}$, at $k_{\ast} = 0.05$ Mpc$^{-1}$,
\begin{equation}
f_{\rm iso}^{2} \equiv
\left. \frac{\Delta_{\mathcal{S}}^{2}}{\Delta_{\mathcal{R}}^{2}}\right\rvert_{k=k_{\ast}}\;.
\label{eq:fiso}
\end{equation}
The corresponding total matter power spectra required in our analysis are computed using the public Boltzmann solver CLASS \citep{Lesgourgues2011}. 
At late cosmic times, however, structure formation becomes non-linear. All relevant modes will have crossed the horizon at this point such that
isocurvature perturbations grow exactly like their adiabatic counterparts (see, e.g., right panel of figure 2 in \cite{Feix2019}). While mildly
non-linear scales are accessible through analytic modeling, the strongly non-linear regime is efficiently approached with the help of $N$-body
simulations, especially when baryonic physics is included. CLASS adopts an implementation of HALOFIT to obtain corrections to the linear power
spectrum \citep{smith_stable_2003, takahashi_revising_2012, mead_accurate_2015}. Since the model and its calibration assume a particular form of
the linear power spectrum, it is unclear how accurate HALOFIT predictions remain in the ALP DM framework, especially since it has been calibrated
against adiabatic fluctuations only. From this point of view, any results that are sensitive to the precise modelling of non-linear scales should
be considered as ballpark estimates. Much more effort involving dedicated simulations is needed to achieve a similar level of precision as currently
established within the standard cosmological model \citep{Smith2019} and will be subject to future investigations. 

\begin{center}
\begin{figure}
\centering
\includegraphics[width=0.6\textwidth]{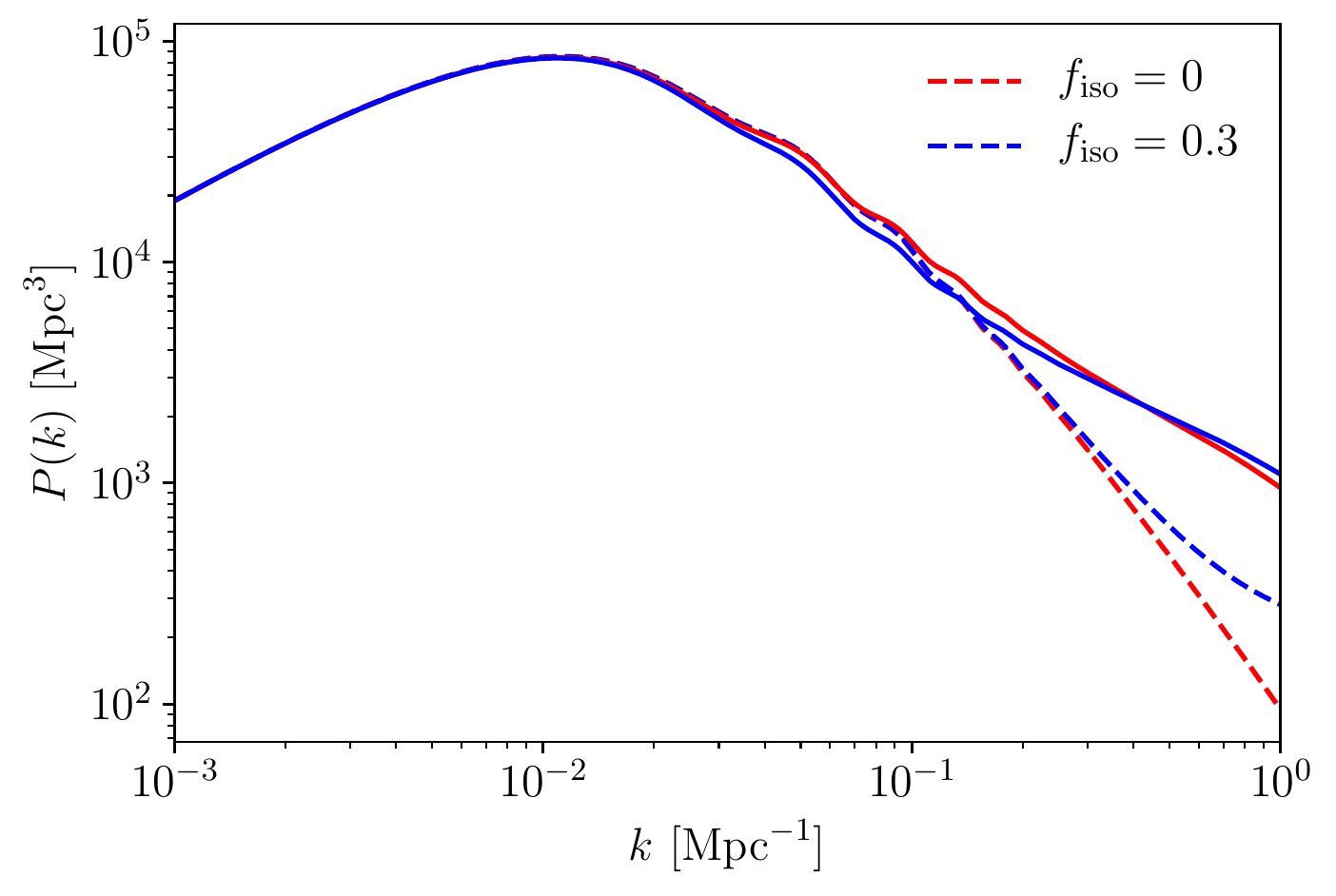}
\caption{Imprint of the ALP DM isocurvature mode generated after inflation on the matter power spectrum $P(k)$ at $z = 0$
(expressed in comoving gauge) for $f_{\rm iso} = 0.3$ (blue) and adiabatic modes only (red). Shown are results for
linear (dashed lines) and non-linear spectra obtained with HALOFIT (solid lines). All power spectra have been evaluated at $z=0$.}

\label{fig:pk}
\end{figure}
\end{center}

\begin{center}
\begin{figure}
\centering
\includegraphics[width=0.485\textwidth]{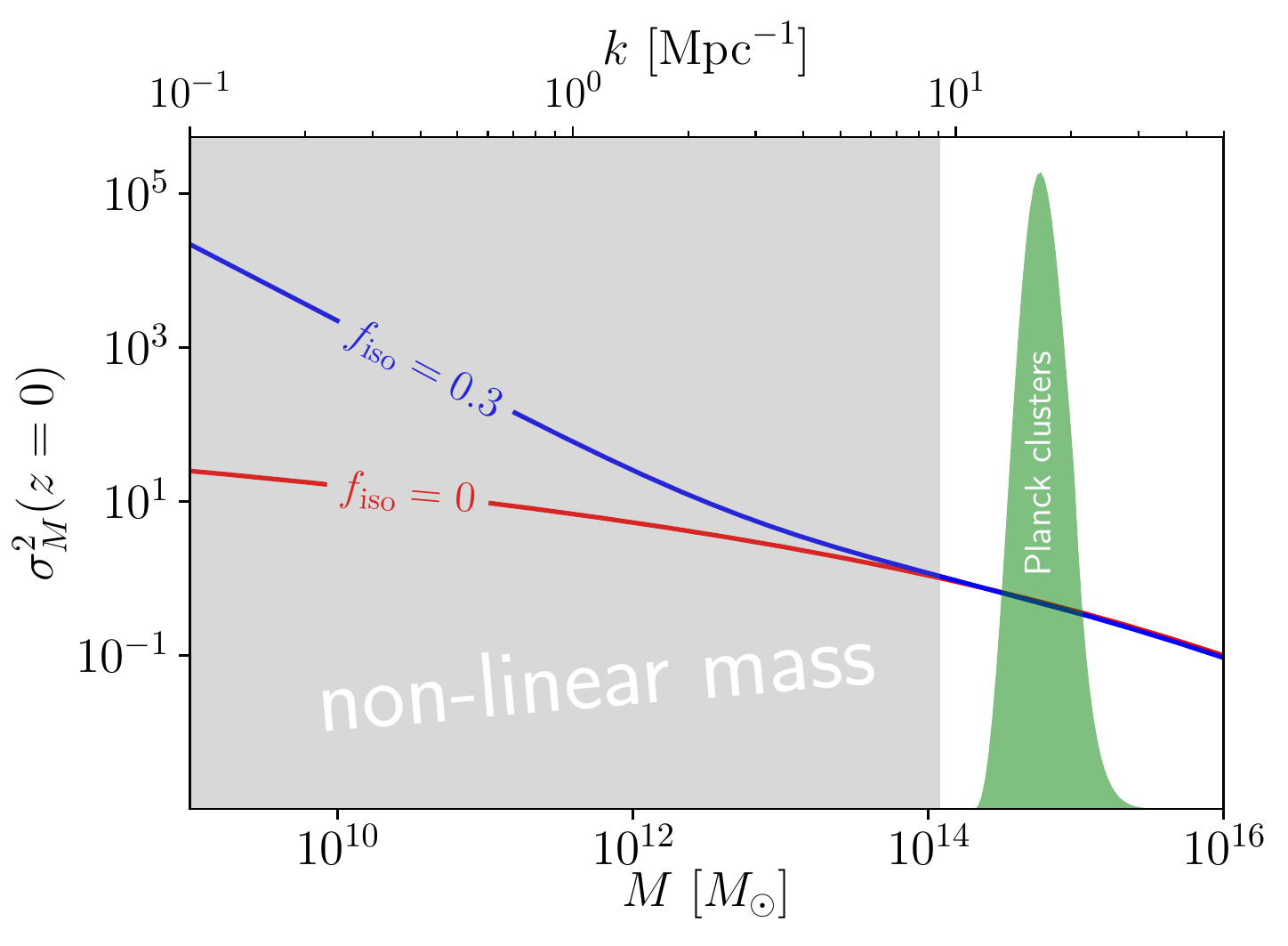}
\hfill
\includegraphics[width=0.485\textwidth]{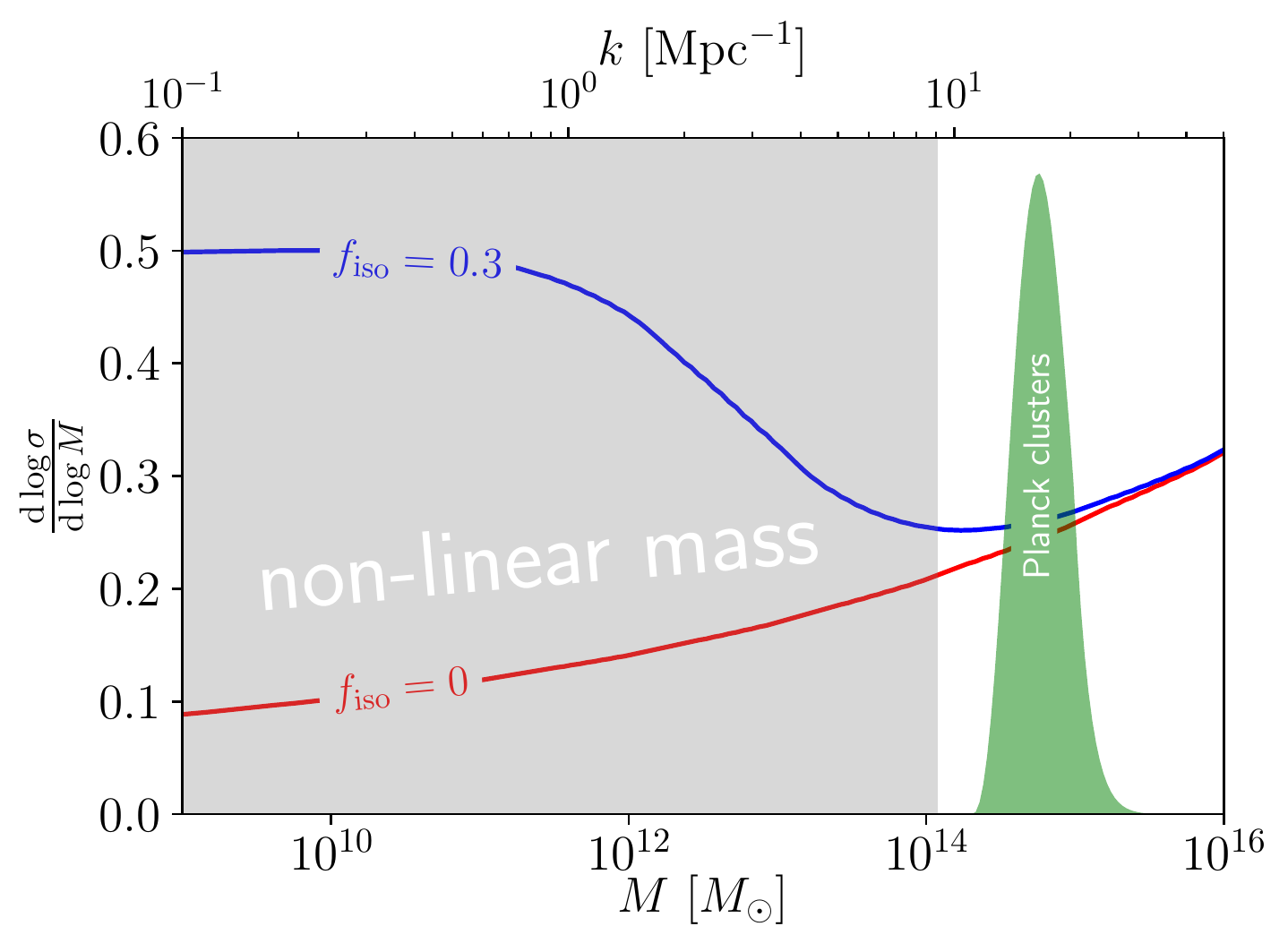}
\caption{Imprint of the ALP DM isocurvature mode generated after inflation on the matter power spectrum variance $\sigma^2(k)$
(left) and its logarithmic derivative (right) at $z = 0$ for $f_{\rm iso} = 0.3$ (blue) and adiabatic modes only (red). All results
adopt the linear power spectrum. The grey region marks the regime of non-linear structure formation while green indicates the mass
range of Planck clusters. The variance is obtained from a top-hat filter with scale $R$ related to the mass through $M = 4\pi\overline{\rho} R^3/3$.
}
\label{fig:sigma}
\end{figure}
\end{center}

In \cref{fig:pk}, we illustrate the matter power spectrum at redshift zero. Blue lines include an additional isocurvature component (assuming $f_{\rm iso} = 0.3$)
and red lines refer to the spectrum with adiabatic fluctuations only. The solid lines correspond to the HALOFIT result while dashed lines represent
the linear spectrum. Non-linear structure formation leads to a redistribution of power through mode coupling, and the additional small-scale power in the isocurvature model accelerates this effect. Therefore the non-linear halofit prediction including isocurvature on intermediate scales can lie below the linear theory line shown in \cref{fig:pk} as additional power is transferred to small scales. In \cref{fig:sigma}, we show the variance $\sigma_{M}^{2}$ of the linear density field and its logarithmic derivative with respect to
$M = 4\pi\overline{\rho} R^3/3$. Here $\bar{\rho}$ is the mean background density and the color scheme is the same as in \cref{fig:pk}. In grey, we show the
region in mass (or scale) that is affected by non-linear gravitational evolution at $z=0$. The green area indicates the effective mass range for clusters in
the Planck sample (see \cref{subsec:clusters}). 

\section{Current constraints from CMB, BAO and clusters}
\label{sec:constraints}

In this section, we revisit the previous CMB constraints on $f_\mathrm{iso}$ from \cite{Feix2019} using the final Planck data release, and add information from
late-time measurements in the form of baryon acoustic oscillations (BAO) and the abundance of galaxy cluster detected by Planck through the Sunyaev Zel'dovich (SZ) effect.
First, we discuss constraints from the primary CMB in \cref{subsec:CMB} before explaining the cluster likelihood in \cref{subsec:clusters} and combining it
with CMB lensing and the SZ cluster data that are both sensitive to the growth of structures at later times in \cref{subsec:combined_constraints}.

\subsection{Primary CMB revisited}
\label{subsec:CMB}

For the constraints presented here, we make use of the latest Planck CMB temperature and polarization measurements \citep{Akrami:2018vks,Aghanim:2018eyx}. The
reconstruction of the CMB lensing signal \citep{Aghanim:2018oex} is sensitive to the amplitude of the density field at later times, and we discuss its effect
together with other late-time probes of large-scale structure in \cref{subsec:combined_constraints}. The CMB measurements are incorporated into the publicly available
\texttt{plik} likelihood code \citep{Aghanim:2019ame}, and all results discussed were obtained by varying cosmological parameters together with all associated
nuisance parameters using the \texttt{MontePython} MCMC sampler \citep{Audren:2012wb,Brinckmann:2018cvx}. Throughout, we assume a standard flat $\Lambda$CDM model
with fixed minimal neutrino masses $\sum m_\nu = 0.06~\mathrm{eV}$ extended by the free isocurvature fraction, $f_\mathrm{iso}$, with a flat prior unless stated otherwise.

We combine the primary CMB data with measurements of the BAO scale from the 6dFGS \cite{Beutler:2011hx}, SDSS-MGS \cite{Ross:2014qpa}, and BOSS DR12 \citep{Alam:2016hwk}
galaxy surveys which help to break geometric degeneracies while making minimal assumptions on the growth of perturbations.

\begin{figure}[tbp]
\centering
\includegraphics[width=.48\textwidth]{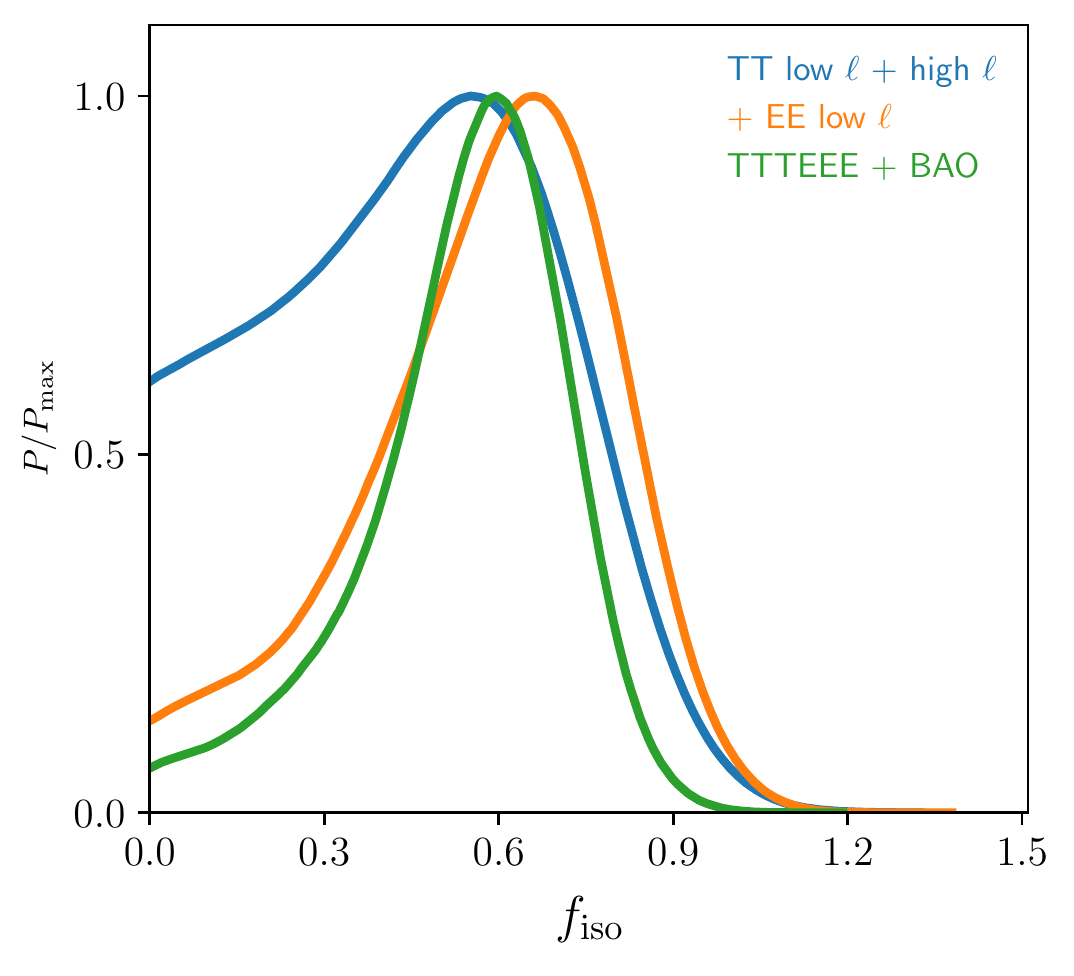} 
\hfill
\includegraphics[width=.48\textwidth]{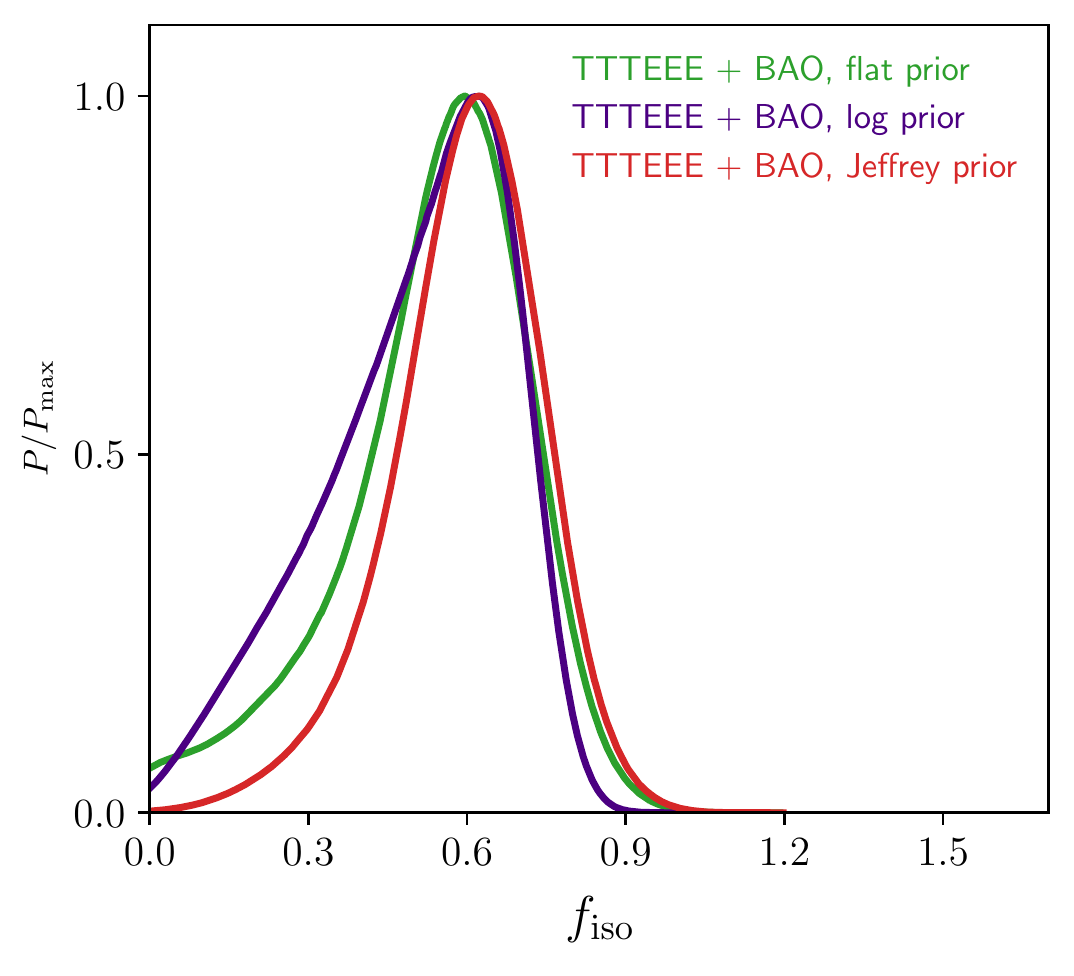}
\caption{\label{fig:fiso_constraint_CMB}
\textbf{Left:} Marginalized posterior distribution of $f_\mathrm{iso}$ from different CMB datasets. The temperature power spectrum (blue) alone is consistent with vanishing
isocurvature contribution, but requires a high value of $\tau \approx 0.1$. Adding EE low-$\ell$ polarization data (orange) enforces a small $\tau$ and leads to a $\sim 3 \sigma$
preference for non-vanishing $f_\mathrm{iso}$. Adding the full polarization spectrum, cross correlations, and BAO information tightens the errors around the best-fit value
$f_\mathrm{iso} = 0.55$. \textbf{Right:} Effect of different priors (described in the text) on the marginalized posterior distribution of $f_\mathrm{iso}$ using the TTTEEE + BAO
dataset. While the width of the distribution changes, the qualitative result is prior-independent.}
\end{figure}

Note that our results from the primary CMB analysis differ from the previous limits on $f_\mathrm{iso}$ obtained by \cite{Feix2019} since we fixed a bug in the MCMC sampler.
The resulting marginalized posteriors for $f_\mathrm{iso}$ using different combinations of CMB datasets are shown in the left panel of \cref{fig:fiso_constraint_CMB}. The
data shows a preference of varying degree for a non-vanishing $f_\mathrm{iso}$. While, due to a strong degeneracy with the optical depth $\tau$, this is very modest for
data from the temperature spectrum alone, adding large-scale EE polarization data enforces $\tau \approx 0.05$, and we find a $2.8 \sigma$ preference for $f_\mathrm{iso} > 0$.
Adding the full polarization EE and TE likelihoods increases the preference to $3.1 \sigma$ since the error bars shrink while the peak of the posterior stays in place.

As a next step, we investigate how much of this formal preference is driven by prior volume effects. The standard analysis uses a flat prior on $f_\mathrm{iso}$, and we consider both a
prior on $\log f_\mathrm{iso}$ that puts more weight on small values, and the least informative Jeffrey prior defined as
\begin{equation}
\label{eq:Jeffrey_prior}
P_J(\theta) \propto \sqrt{ \det \mathcal F(\theta)} \: ,
\end{equation}
where the Fisher matrix $\mathcal F$ derived for a CMB experiment is described in \cite{Feix2019}. The prior scales approximately like $P_J(f_\mathrm{iso}) \propto f_\mathrm{iso}$
and, therefore, puts more weight on larger $f_\mathrm{iso}$ values. The results from using these different priors are shown in the right panel of \cref{fig:fiso_constraint_CMB}.
While the width of the distribution slightly changes, the overall effect is modest. This is also reflected in the $\chi^2$-value of the best-fit isocurvature models, where the
introduction of $f_\mathrm{iso}$ improves the fit by $\Delta \chi^2 = -6.7$ over $\Lambda$CDM for the TTTEEE + BAO dataset. The marginalized results for $f_\mathrm{iso}$,
together with the associated changes in $\chi^2$ are summarized in \cref{tab:chi2}.

\begin{figure}
\centering
\includegraphics[width=1\textwidth]{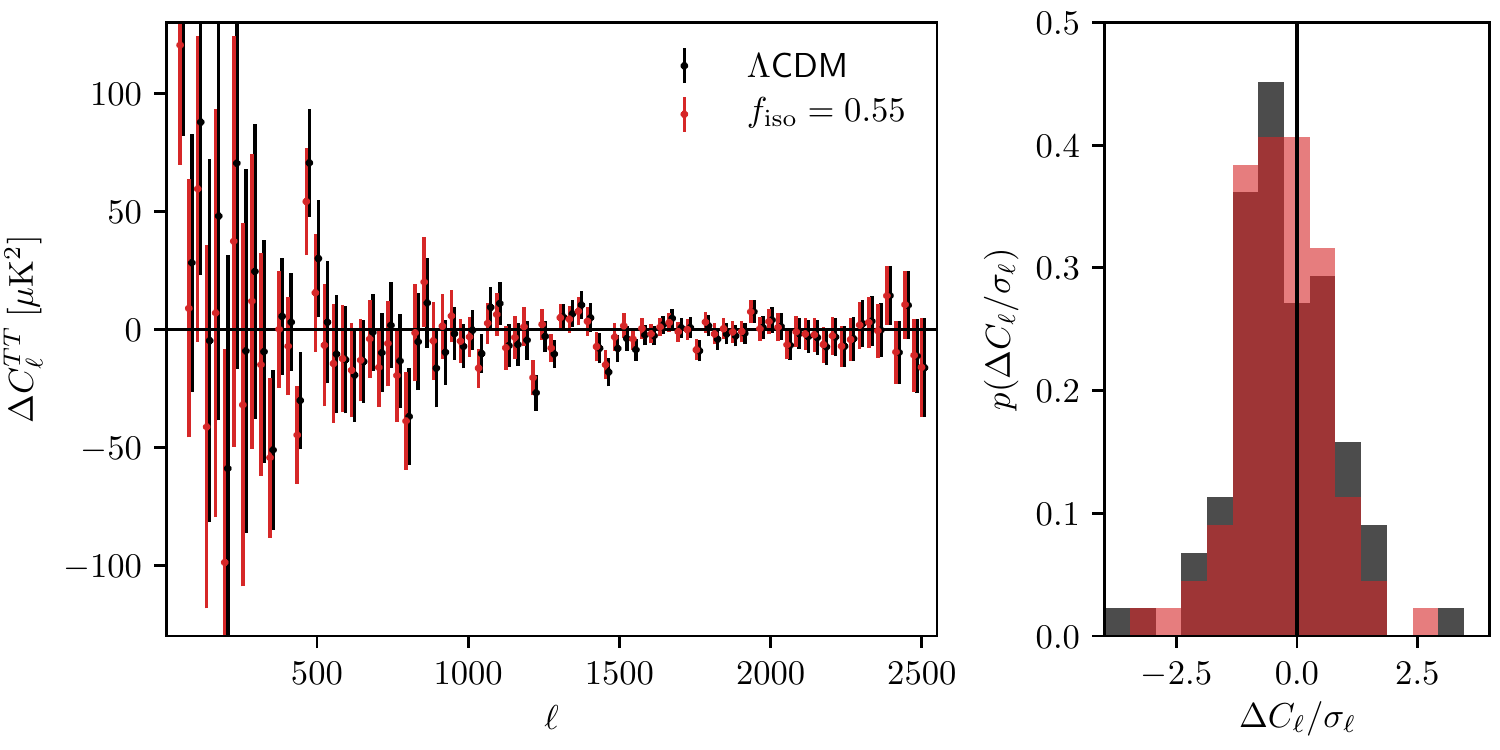} 
\caption{\label{fig:CMB_residuals} \textbf{Left:} Residuals of the binned Planck temperature spectrum for the best-fit $\Lambda$CDM (black) and $f_\mathrm{iso}$ (red) models.
The improved fit to the temperature spectrum is primarily due to low multipoles and the oscillatory feature that is also responsible for the $A_L$
anomaly around $\ell \approx 1000 - 1800$.
\textbf{Right:} Histogram of the respective normalized best-fit residuals for the binned temperature power spectrum. While the $\Lambda$CDM residual distribution is broader, a Kolmogorov-Smirnov test as described in the text cannot distinguish between both distributions. }
\end{figure}

The residuals of the binned Planck temperature spectrum for the best-fit $\Lambda$CDM and isocurvature models are shown in \cref{fig:CMB_residuals}. The
improvement in this case partly stems from the ability to fit two well-known features in the CMB data: Planck observes a slight lack of power on very large scales and an oscillatory
feature in the residuals often associated with the fudge parameter $A_L$ since it can be described by scaling up the effect of gravitational lensing on the temperature
spectrum\footnote{We emphasize that the additional lensing effect is not detected by the CMB lensing likelihood itself as is also discussed in \cite{Aghanim:2018eyx} and \cref{subsec:combined_constraints}}. Both these effects are extensively discussed in \cite{Aghanim:2018eyx} and are consistent with a statistical fluctuation in
$\Lambda$CDM. Just considering the primary CMB data, the isocurvature component can mimic the behavior of $A_L$ by boosting the matter power spectrum as seen in \cref{fig:pk}
without changing the standard cosmological parameters. This leads to a stronger CMB lensing effect compared to $\Lambda$CDM. 
The preferred amplitude for $A_L$ is largest for the temperature power spectrum alone and decreases as more polarization data are added \cite{Aghanim:2018eyx} or when a larger
sky fraction is used \cite{Efstathiou2019}, which is consistent with the behavior expected from a statistical fluctuation. Although the preference for $f_\mathrm{iso}$ is linked
to the same multipoles responsible for the $A_L$ anomaly, the behavior is somewhat different and the preferred value is not reduced when adding polarization data.

The distribution of the
normalized temperature power spectrum residuals (shown on the right-hand side of \cref{fig:CMB_residuals}) indicates again a slight preference for the isocurvature model,
which is reflected in the broader histogram of $\Lambda$CDM residuals. For a quantitative comparison, we apply a Kolmogorov-Smirnov-test to see if the two samples of residuals
could have originated from the same distribution. This yields a score of $S=0.121$, which is less than the critical value of $D = 0.189$ for $n=83$ binned data points, choosing
the size of the test as $\alpha=0.1$\footnote{Then, $c(\alpha) = 1.22$ and the critical value is approximately given by $D = c(\alpha)\sqrt{2/n}$}.
The resulting probability that both samples are drawn from the same distribution is $\sim 58 \%$, so the test is indecisive and does not find a significant difference between the
residual distributions of the two models.

\begin{figure}
\centering
\includegraphics[width=0.7\textwidth]{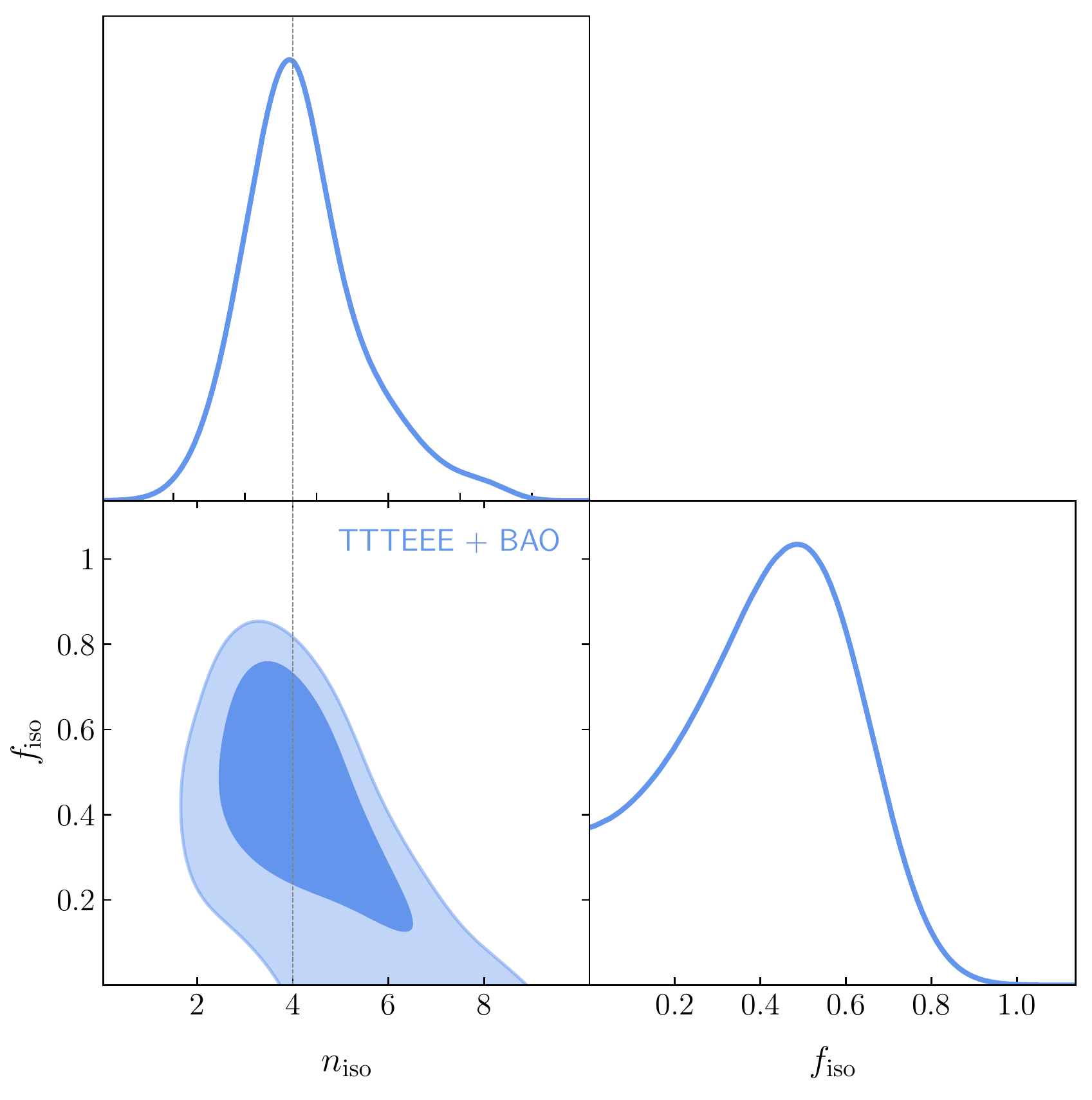} 
\caption{\label{fig:fcdi_ncdi_triangle} Marginalized $68 \%$ and $95 \%$ posterior contours for the TTTEEE + BAO dataset, varying both the amplitude
$f_\mathrm{iso}$ and the spectral tilt $n_\mathrm{iso}$ of the isocurvature spectrum.}
\end{figure}

Note that the spectral index of the isocurvature componented predicted in our scenario is fixed by the white-noise shape due to causality: from \cref{Eq:Delta} we see that $n_{\rm iso} = 4$ in our model, where the spectral index is defined as usuall by $\Delta^2(k) \propto k^{n_{\rm iso}-1}$. Hence, in our default analysis presented sofar the isocurvature spectral index is not a free parameter and we fix $n_{\rm iso} = 4$. For illustrative purposes, let us now relax this assumption and introduce $n_{\rm iso}$ as an additional free parameter. The results of this analysis are shown in 
\cref{fig:fcdi_ncdi_triangle}. Interestingly, the improvement in the CMB fit
is specifically linked to a spectrum with $n_\mathrm{iso} \approx 4$. If the isocurvature spectrum has a smaller tilt, it causes signatures in the CMB spectra on large scales
while a blue spectrum introduces additional power at very high $\ell$. This is particularly interesting since, as mentioned above, $n_\mathrm{iso} = 4$ is a specific prediction for the isocurvature
component produced by axion miniclusters. While the isocurvature model improves the fit to the primary CMB spectra, it also results in larger amplitudes $\sigma_8$ of matter
fluctuations in the late universe due to additional $f_\mathrm{iso}$-contributions on small scales. In \cref{subsec:combined_constraints}, we discuss this issue in more detail
when combining CMB and LSS measurements.

\subsection{Planck SZ clusters}
\label{subsec:clusters}

The cosmological Planck cluster sample consists of 438 massive objects detected through their SZ imprint in the CMB maps. We follow the modelling outlined in the original analysis \cite{PlanckSZ_13, PlanckSZ_15}, where the catalog is binned in both redshift $z$ and signal-to-noise of the SZ detection $q$. Since the Planck SZ sample consists of rare objects from the high-mass tail of the halo mass function, the likelihood is well approximated by a Poisson distribution, which we correct for sample variance effects \cite{Takada2014}. The likelihood depends on the expected number of cluster counts per bin $(\Delta z_i, \Delta q_j)$, which can be written as
\begin{equation}
N(\Delta z_i, \Delta q_j) = \int_{\Delta z_i} \mathrm d z \int_{\Delta q_j} \mathrm d q \frac{\mathrm d n}{\mathrm d z \mathrm d q} \: ,
\end{equation}
where the density of clusters is a function of signal-to-noise,
\begin{equation}
\label{eq:dndq}
\frac{\mathrm d n}{\mathrm d z \mathrm d q} = \int \mathrm d M_{500c} \frac{\mathrm d n(M_{500c},z)}{\mathrm d M_{500c}} p(q | M_{500c}, z) \: ,
\end{equation}
and the Planck clusters are defined as spherical overdensities up to the radius $R_{500}$ where the mean density inside is equal to $500$ times the critical density $\rho_c$, {$M_{500c} = 4/3 \pi R_{500}^3 500 \rho_c(z)$.

To constrain cosmology with the cluster abundance, we need to specify the mass function $\mathrm d n/ \mathrm d M_{500c}$ and the observable-mass relation $p(q|M_{500c}, z)$. Without $N$-body simulations to calibrate the cluster abundance, we follow a simplified rescaling procedure based on the linear change in the variance of the density field $\sigma_M^2$ alone. We assume that the relative effect of $f_\mathrm{iso}$ on the cluster abundance is captured by writing 
\begin{equation}
\left . \frac{\mathrm d n^\mathrm{T}}{\mathrm d M_{500c}} \right \rvert_\mathrm{iso}= 
\frac{\left . \frac{\mathrm d n^\mathrm{PS}}{ \mathrm d M} \right \rvert_\mathrm{iso}}{\left . \frac{\mathrm d n^\mathrm{PS}}{ \mathrm d M} \right \rvert_\mathrm{fid}} 
\left .\frac{\mathrm d n^\mathrm{T}}{\mathrm d M_{500c}} \right \rvert_\mathrm{fid} \: ,
\end{equation}
where the analytical Press-Schechter mass functions $n^\mathrm{PS}$ \citep{Press1974} are evaluated using the variance of the density field including isocurvature (denoted by the subscript ``iso'') or the fiducial variance in $\Lambda$CDM (denoted by ``fid''), and $n^\mathrm{T}$ is the Tinker mass function \citep{Tinker2008} calibrated to $\Lambda$CDM $N$-body simulations. This ensures that the standard $\Lambda$CDM results are recovered for $f_\mathrm{iso} \rightarrow 0$. Since the Press-Schechter mass function is defined for virialized halo masses, we rescale it to $M_{500c}$, assuming NFW density profiles and using the procedure outlined in the appendix of \cite{Hu2003} with the empirical halo mass-concentration relation from \cite{Dutton2014}.

The second ingredient, the probability distribution $p(q | M_{500c}, z)$ in \cref{eq:dndq}, is governed by baryonic physics and, therefore, unaffected by $f_\mathrm{iso}$. Cosmological constraints from cluster abundance and potential tensions with the primary CMB depend critically on the mean mass calibration of the sample, expressed primarily by the mass bias parameter $(1-b_H)$ that accounts for an offset between hydrostatic and true cluster masses. It has to be determined by external measurements (see the extensive discussion in \cite{PlanckSZ_13, PlanckSZ_15, Zubeldia2019}), and for this work, we follow the recent analysis using the lensing imprint of the stacked Planck SZ clusters on the CMB performed by \cite{Zubeldia2019}, leading to a Gaussian prior $(1-b_H) = 0.71 \pm 0.10$. All other nuisance parameters of $p(q|M_{500c}, z)$ are varied with the same priors as in \cite{PlanckSZ_15, Zubeldia2019}, and we refer to those papers for details. Since the abundance of galaxy clusters is mostly sensitive to the total amplitude of matter fluctuations set by $\Omega_{\rm m}$ and the standard deviation of the density field smoothed at 8 Mpc/$h$, $\sigma_8$, we combine the cluster likelihood either with the CMB data, or with a Gaussian prior on the baryon density from big-bang nucleosynthesis $\Omega_{\rm b} h^2 = (2.224 \pm 0.046) \times 10^{-2}$ \cite{Cooke2014} and the combined BAO measurements mentioned in \cref{subsec:CMB} to constrain the other cosmological parameters. For simplicity, we present results from clusters in terms of the main degeneracy direction $S_8 = \sigma_8 \sqrt{\Omega_{\rm m} / 0.3}$.

\begin{figure}
\centering
\includegraphics[width=.48\textwidth]{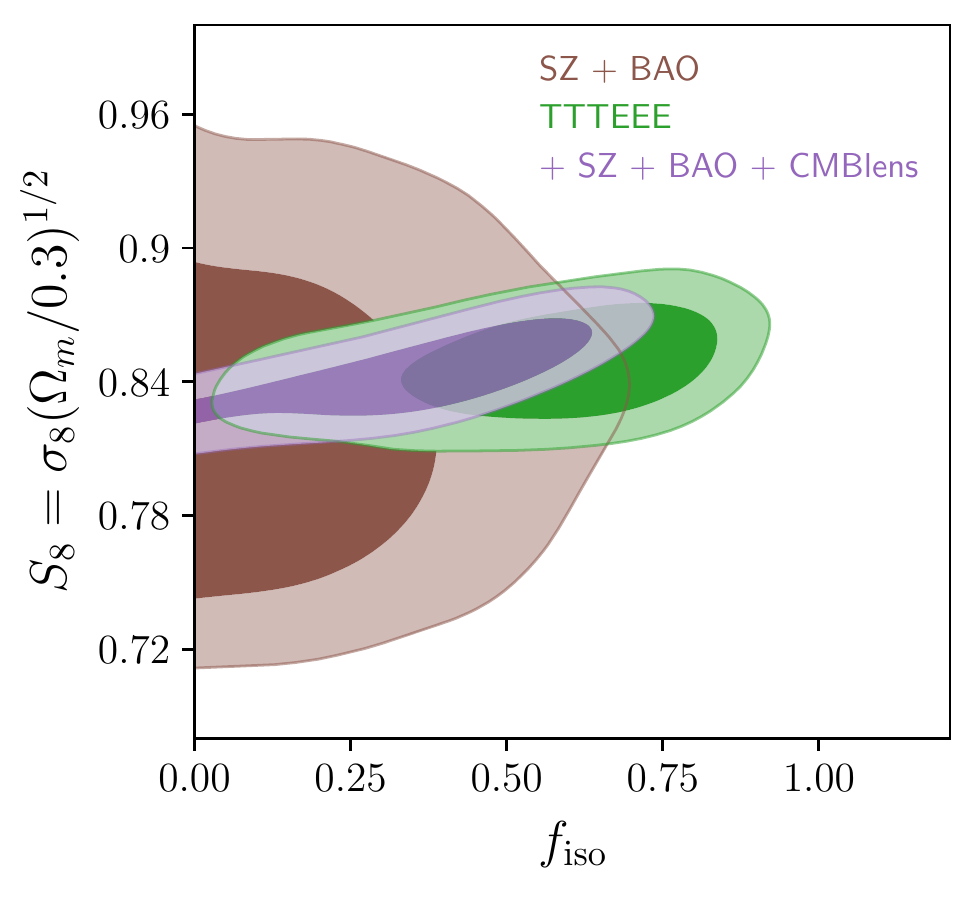} 
\hfill
\includegraphics[width=.48\textwidth]{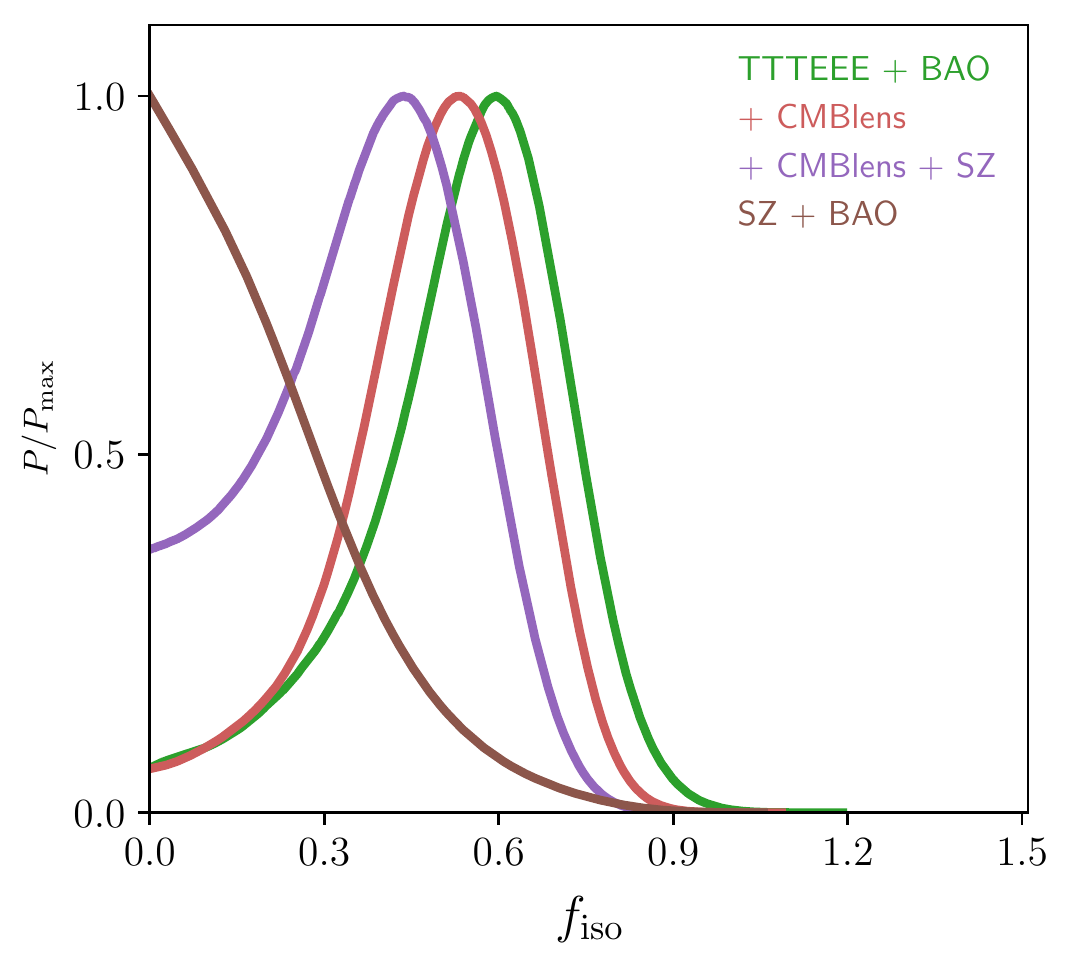}
\caption{\label{fig:fiso_constraint_CMB_LSS} \textbf{Left:} Marginalized $68 \%$ and $95 \%$ contours in the $f_\mathrm{iso} - S_8$ plane for primary CMB (green), Planck SZ Clusters (SZ + BAO, brown) and their combination including CMB lensing (purple). Constraints from individual datasets are mutually consistent and can be safely combined. \textbf{Right:} Marginalized posterior distribution of $f_\mathrm{iso}$ for various combinations of probes. The SZ+BAO data (brown) are consistent with $f_\mathrm{iso}=0$, and consequently the preference from primary CMB+BAO data alone (green) is shifted towards smaller values as more LSS datasets are added.}
\end{figure}

\subsection{Combined constraints from early and late times}
\label{subsec:combined_constraints}

As mentioned in \cref{subsec:CMB}, one signature of isocurvature perturbations is a larger amplitude of the matter power spectrum that can be detected by large-scale structure experiments.
Now we combine the primary CMB data with measurements of the CMB lensing signal and the Planck SZ cluster likelihood described in the previous \cref{subsec:clusters}.

\begin{table}
\centering
\renewcommand{\arraystretch}{1.4}
\begin{tabular}{l c c r}
Dataset &   $f_\mathrm{iso}$ & $95 \%$ upper limit &$\Delta \chi^2$ \\
\hline
\hline
TT 							&  $0.46^{+0.27}_{-0.26}$ 	& $<0.82$ 	& $-2.8$			\\ 
TT + EE low $\ell$  				& $0.58^{+0.26}_{-0.16}$  	& $<0.93$		& $-3.2$ 			\\
TTTEEE 						& $0.57^{+0.21}_{-0.13}$ 		& $<0.88$		& $-6.2$			\\
TTTEEE + BAO 				& $0.55^{+0.20}_{-0.12}$ 		& $<0.85$		& $-6.7$ 			\\
TTTEEE + BAO + CMBlens 		& $0.50^{+0.17}_{-0.12}$ 		& $<0.78$		& $-3.5$ 			\\
SZ + BAO 					& $< 0.27$ 				& $<0.53$		& $0$			\\
TTTEEE + BAO + CMBlens + SZ 	& $0.37^{+0.21}_{-0.15}$ 		& $<0.64$		& $-1.2$			\\
\hline
\end{tabular}
\caption{\label{tab:chi2} Overview of all data combinations employed in the analysis together with the respective mean $f_\mathrm{iso}$ and marginalized $68\%$ confidence intervals and the $95 \%$ upper limit. The $\Delta \chi^2$ is calculated as the difference between best-fit $\Lambda$CDM and isocurvature models. All results shown here are derived assuming a flat prior on $f_\mathrm{iso}$.}
\end{table}

The number of detected Planck SZ clusters is slightly low given the CMB best-fit $\Lambda$CDM cosmology, but both datasets are compatible with the cluster mass calibration used here \cite{PlanckSZ_15, Zubeldia2019}. However, any discrepancy already present under the assumption of $\Lambda$CDM becomes more pronounced in the extended $f_\mathrm{iso}$ cosmology since the standard $\Lambda$CDM parameters inferred from the CMB do not change much whereas the additional white-noise isocurvature component leads to larger values of $S_8$ compared to the standard cosmological model. We, therefore, start by considering separate constraints in the $f_\mathrm{iso} - S_8$ plane from primary CMB and SZ+BAO. These are presented in the left panel of \cref{fig:fiso_constraint_CMB_LSS}. Although both posteriors are compatible, the SZ+BAO likelihood does not show a preference for $f_\mathrm{iso} > 0$. Joint posteriors for various data combinations are shown in the right panel of \cref{fig:fiso_constraint_CMB_LSS}. The CMB lensing likelihood shows the same tendency to lower the isocurvature level preferred by the primary CMB. Adding both CMB lensing and the SZ cluster abundance likelihoods to the primary CMB lowers the marginalized posterior constraint from $f_\mathrm{iso} = 0.55^{+0.20}_{-0.12}$ to $f_\mathrm{iso} = 0.37^{+0.21}_{-0.15}$ which is consistent with zero at the $95 \%$ confidence level.

\Cref{tab:chi2} gives a summary of marginalized $f_\mathrm{iso}$-posteriors together with the resulting improvement in $\chi^2$ over the fiducial best-fit $\Lambda$CDM cosmology for all data combinations. Although the significance for $f_\mathrm{iso}$ increases with additional primary CMB and BAO data, LSS probes show no indication of an isocurvature signature. For the combination of all datasets, the improvement is very modest, $\Delta \chi^2 = -1.2$, and even without performing a full Bayesian model comparison, there is no significant preference for the extended model over $\Lambda$CDM.

While the specific axion model explored here fits a real feature present in the primary CMB temperature and polarization data (as indicated by the improvement in $\chi^2$), we caution to interpret this as a signature of isocurvature perturbations. Such an interpretation is disfavored by current LSS data. Until the situation is resolved by additional measurements, we assume that either a statistical fluctuation or another unknown effect in the CMB causes the preference for $f_\mathrm{iso}$. This in turn leads to weaker upper bounds on the model than expected in an idealised forecast as performed in \cite{Feix2019}. The preferred values for $f_\mathrm{iso}$ suggested by CMB data are also in tension with limits derived from Ly-$\alpha$ and re-ionization in \citep{Irsic2019}.

\begin{figure}
  \includegraphics[width=0.48\textwidth]{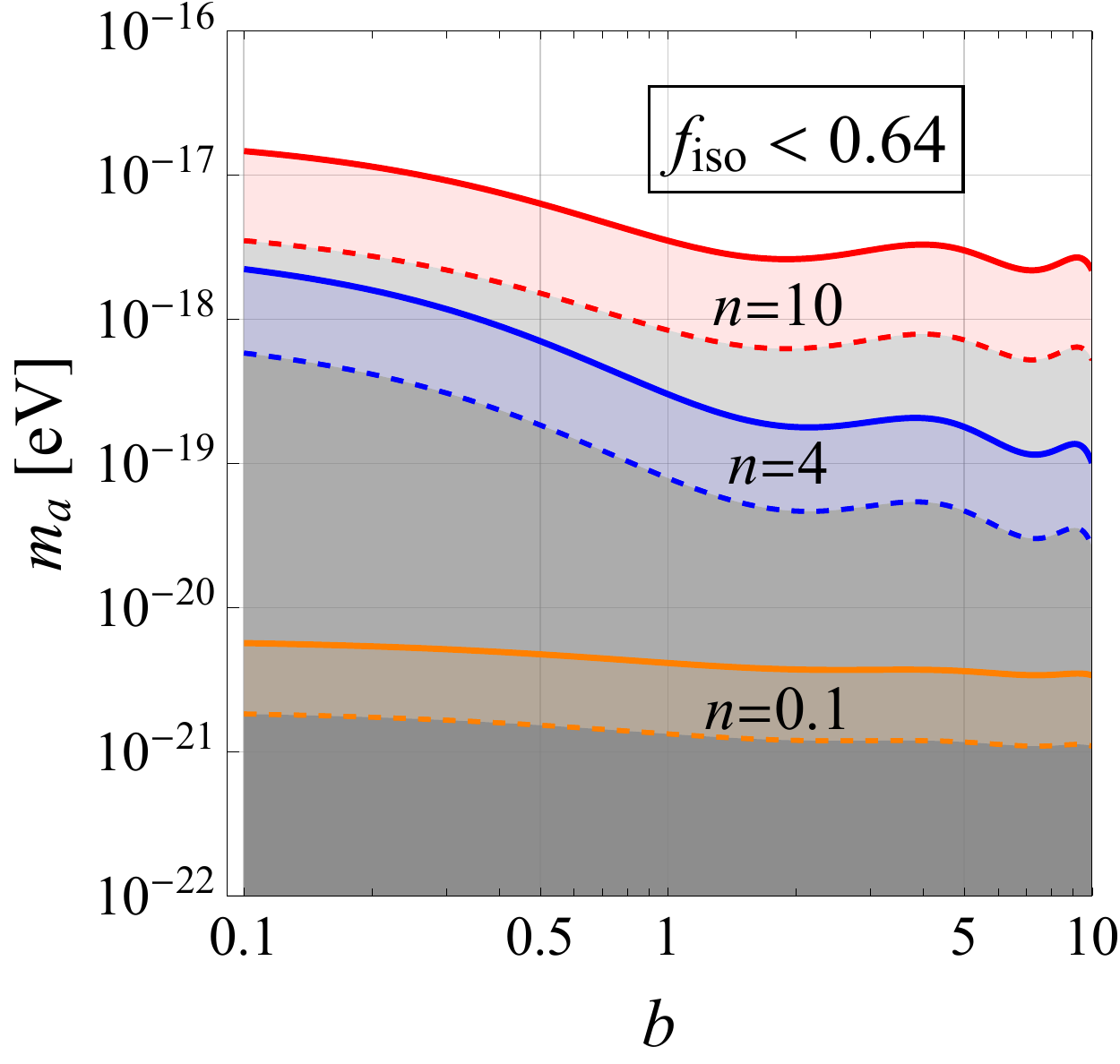}
  \hfill
  \includegraphics[width=0.48\textwidth]{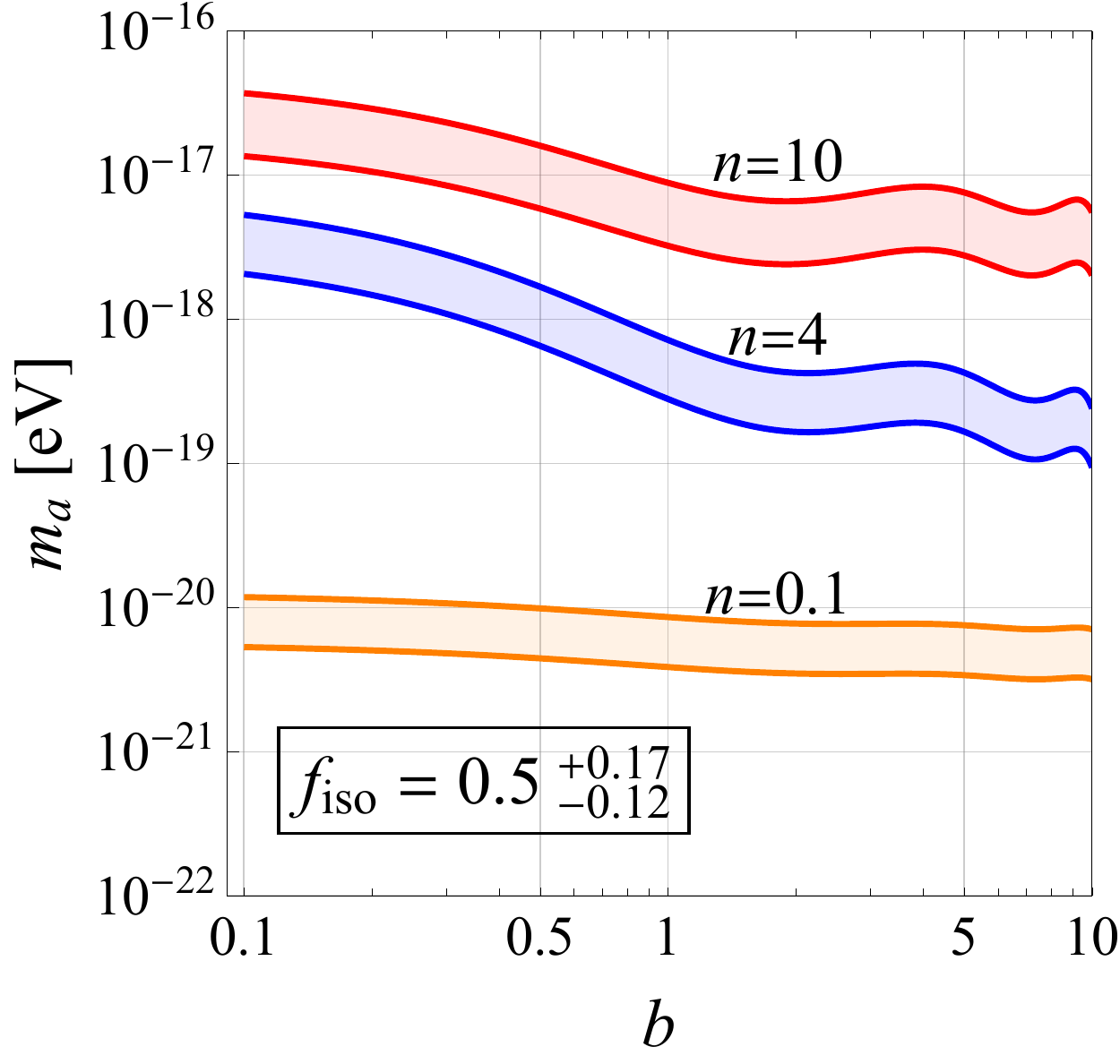}
  \caption{\textbf{Left:} Conservative constraints on the
    zero-temperature ALP mass $m_{a}$ from CMB TTTEEEE + BAO + CMBlens
    + SZ. Solid curves correspond to bounds and bands between solid
    and dashed curves indicate a factor 5 uncertainty. The region
    below the curves is disfavored. \textbf{Right:} ALP masses which
    could explain the non-zero value of $f_{\rm iso}$ preferred by CMB
    data. In both plots, we show results for different assumptions on
    the ALP mass-temperature dependence as given in
    \cref{Eq:MassParameterization}, and ALPs are assumed to provide
    all DM. The case $n=0.1$ is numerically very similar to the case
    of a temperature-independent ALP mass with $n=0$.}
\label{fig:mass_bound_current}
\end{figure}

Hence we take a conservative approach and consider our results as an
upper limit of $f_{\rm iso}$. Following \cite{Feix2019}, this limit
can be used to constrain the ALP mass, assuming that all DM was
created by the post-inflationary vacuum misalignment mechanism. The
results are shown in the left panel of \cref{fig:mass_bound_current}
for the 95\%~CL limit from the combined CMB+BAO+SZ analysis, $f_{\rm
  iso} < 0.64$.  We see that these observational data exclude
zero-temperature ALP masses smaller than $10^{-21}$~eV in the case of
a weak temperature dependence of the ALP mass ($n\lesssim 0.1$). Note
that for such small values of $n$ the ALP mass is nearly temperature
independent and the results shown in the plots with $n=0.1$ are
numerically very similar to the case of $n=0$. For a stronger
temperature dependence, the lower bound on the ALP mass can become as
high as $10^{-17}$~eV, excluding relevant regions of parameter space
beyond current bounds such as \cref{eq:bounds}. For illustrative
purposes, the right panel of \cref{fig:mass_bound_current} shows
values of ALP masses that could explain the isocurvature component
preferred by CMB data, i.e. $f_{\rm iso} = 0.5_{-0.12}^{+0.17}$.

Let us comment on the crucial assumption of post-inflationary symmetry
breaking, which for small ALP masses requires a somewhat high scale of
inflation. On the one hand, this offers the additional potential
signature of tensor fluctuations in the observable range. However, on
the other hand, for masses $m_a \sim 10^{-21}$~eV and $n \lesssim
0.1$, relatively high values of $f_a$ are required to match the total
DM relic abundance, $f_a \sim 5\times 10^{16}$~GeV, see left panel of
Fig.~1 in Ref.~\cite{Feix2019}. Hence, in this regime it becomes
difficult to reconcile this scenario with simple inflationary
models. First, we note that more uncoventional inflationary models may
be still consistent with such a high scale of inflation. Second, in
this regime, other astrophysical constrains on ALP DM apply as well
\cite{Marsh2018}. For $n\gtrsim 4$, constraints on the scale of
inflation are somewhat relaxed and the requirements for our bounds to
apply are consistent even with simple single-field inflationary
models, see Sec.~2.4 of Ref.~\cite{Feix2019} for discussion and
further references.

\section{Future prospects from large-scale structure probes}
\label{sec:lss}

In this section, we investigate future probes of the LSS and their sensitivity to isocurvature perturbations induced by ALPs. We will focus on
a photometric survey such as Euclid in combination with the lensing signal of a stage-IV CMB survey. In \cref{sec:cosmic_shear,sec:galaxy_clustering,sec:CMBlens},
we will review the different probes. \Cref{sec:lss-sens} will present possible isocurvature constraints and associated bounds on the ALP mass in
the non-linear description for different combination of probes.

\subsection{Cosmic shear}
\label{sec:cosmic_shear}
Bundles of light rays traveling through the LSS get deformed due to perturbed gravitational potentials \citep[see, e.g.,][for reviews]{bartelmann_weak_2001,
hoekstra_weak_2008}. To first order, the weak lensing effect can be described by a line-of-sight integral of the scalar metric perturbation $\Phi$. The lensing
potential is defined as
\begin{equation}\label{eq:lensingpot}
\psi_{i} = 2\int_{0}^{\chi_H}\mathrm{d}\chi W_{\psi_{i}}(\chi) \Phi\;,
\end{equation}
where $\chi_{H} = c/H$ is the Hubble radius and the index $i$ denotes the tomographic bin. Furthermore, the lensing weight function is
\begin{equation}
W_{\psi_i}(\chi) = \frac{G_i(\chi)}{a\chi}\;,
\end{equation}
which includes the tomographic lensing efficiency function
\begin{equation}
G_i(\chi) = 
\int _{\mathrm{min}(\chi,\chi_i)}^{\chi_{i+1}}\mathrm{d}\chi'p(\chi')\frac{\mathrm{d}z}{\mathrm{d}\chi'}\left(1-\frac{\chi}{\chi'}\right)\;.
\end{equation}
The Jacobi determinant $\mathrm{d}z/\mathrm{d}\chi' = H(\chi')/c$ arises from the transformation of the galaxy redshift distribution $p(z)\mathrm{d}z$
which is modeled as \citep{laureijs_euclid_2011}
\begin{equation}
p(z)\mathrm{d}z \propto z^2\exp\left[-\left(\frac{z}{z_0}\right)^\beta\right]\;,
\end{equation}
where $z_0\approx 0.9$ and $\beta = 3/2$. Finally, the lensing potential's angular power spectrum
in the Limber approximation \citep{limber_analysis_1954} is given by
\begin{equation}\label{eq:cosmic_shear_power}
C_{\psi_i\psi_j}(\ell) = 
\int_0^{\chi_H} \frac{\mathrm{d}\chi}{\chi^2}W_{\psi_i}(\chi)W_{\psi_j}(\chi)P_{\Phi}(\ell'/\chi,\chi)\;.
\end{equation}
Note that we defined
$\ell^\prime = \ell + 1/2$, which we will use for the remainder of this section.

The noise contribution of observed lensing spectra is Poissonian shape noise due to the finite number of galaxies in each bin.
The estimator of the lensing signal is then given by
\begin{equation}
\hat{C}_{\psi_i\psi_j} = C_{\psi_i\psi_j} + \sigma_\epsilon^2\frac{n_\mathrm{bin}}{4\bar{n}}\ell^4\delta_{ij},
\end{equation}
where the intrinsic ellipticity dispersion $\sigma_\epsilon = 0.3$, $n_\mathrm{bin}$ denotes the number of tomographic bins,
and $\bar{n}$ is the mean number density of galaxies.

\subsection{Galaxy clustering}
\label{sec:galaxy_clustering}
Complementary to cosmic shear, galaxy clustering \citep[e.g.][]{baumgart_fourier_1991, feldman_power-spectrum_1994, heavens_spherical_1995,raccanelli_cosmological_2016} measures the statistics of the density contrast, $\delta$, and thus directly the matter power spectrum. However, galaxies are biased tracers of the density field \citep[][for a review]{desjacques_large-scale_2018}. Quite generally, we will, therefore, write $\delta(k,z)b(k,z) = \delta_g(k,z)$. In complete analogy to cosmic shear, the tomographic angular power spectrum is given by
\begin{equation}
C_{g_ig_j}(\ell) = \int_0^{\chi_H} \frac{\mathrm{d}\chi}{\chi^2}W_{g_i}(\ell'/\chi,\chi)W_{g_j}(\ell'/\chi,\chi)P_\delta(\ell'/\chi,\chi),
\end{equation}
where $P_\delta$ is the matter power spectrum.
The galaxy weight function is defined as
\begin{equation}
W_{g_i}(\ell/\chi,\chi) = 
    \frac{H(\chi)}{c}b(\ell/\chi,\chi)p(\chi) \  \mathrm{if} \ \chi\in [\chi_i,\chi_{i+1}).
\end{equation}
We assume no correlations between different tomographic bins since for the number of bins used in our setting, the photometric redshift error is smaller than the bin width. In \citep{bailoni_improving_2017}, these assumptions where discussed quantitatively in great detail for a spectroscopic survey. 
For the galaxy bias, we assume a simple linear model \citep{ferraro_wise_2015}:
\begin{equation}
b_i(\chi) = b_i(1+z(\chi))\;,
\end{equation}
where $b_i$ is the bias parameter for each redshift bin.
Again, the observed spectrum includes a shot-noise component,
\begin{equation}
\hat{C}_{g_ig_j} = C_{g_ig_j} + \frac{n_\mathrm{bin}}{\bar{n}}\delta_{ij}\;.
\end{equation}

\begin{center}
\begin{figure}
\includegraphics[width = 0.485\textwidth]{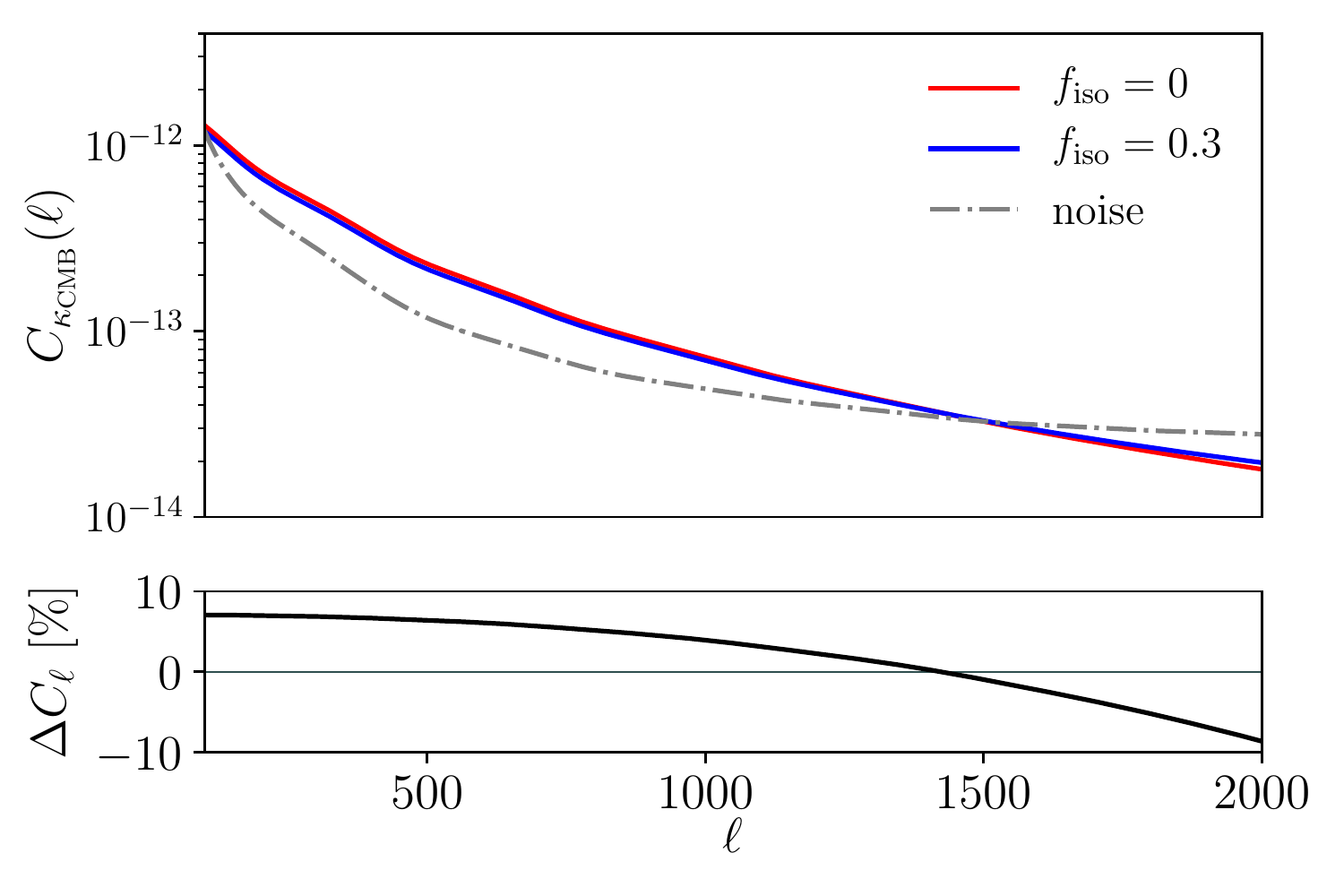}
\hfill
\includegraphics[width = 0.485\textwidth]{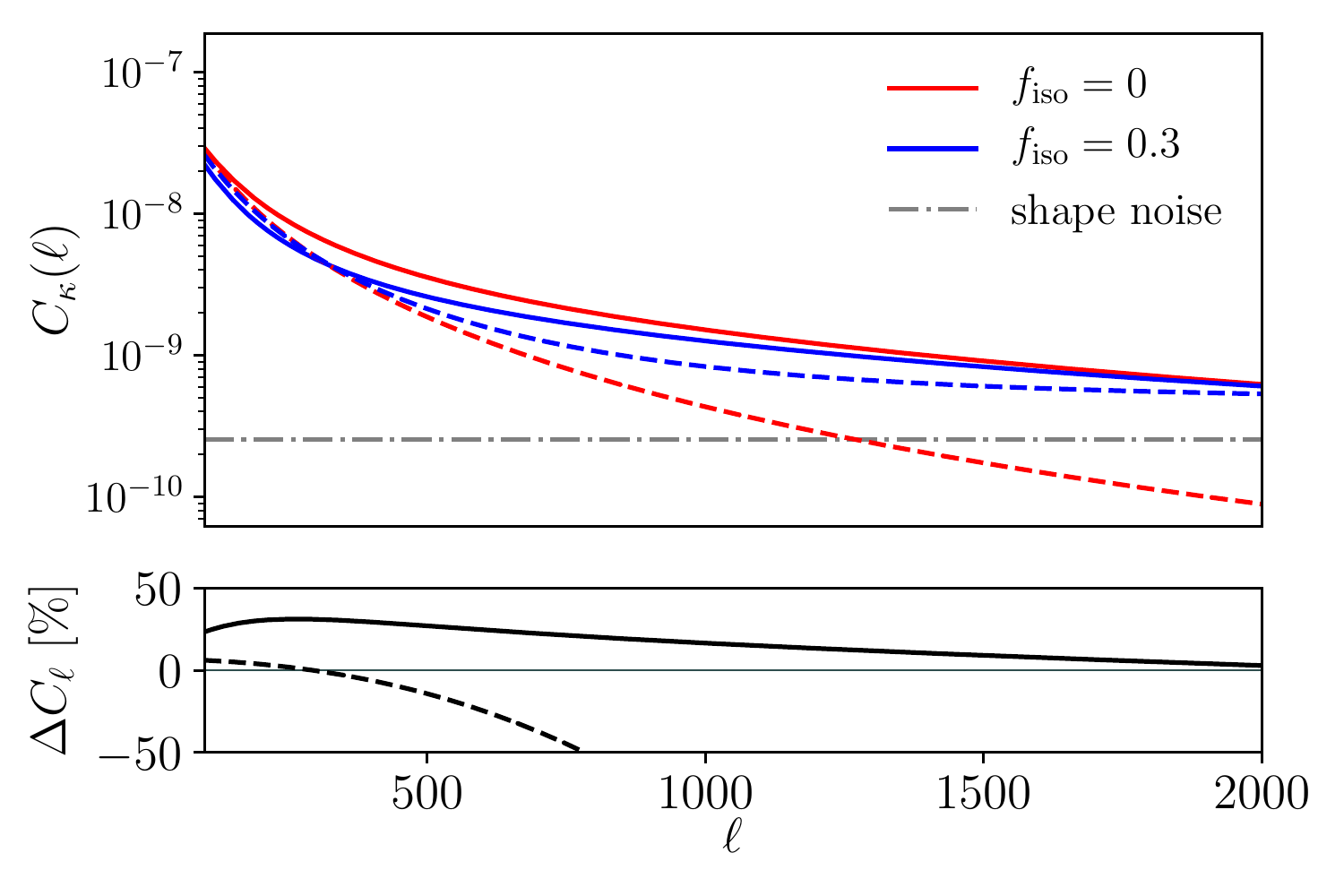}
\caption{\textbf{Left:} The CMB lensing angular power spectrum. Since the largest contributions come from higher redshifts and large scales,
the effect of the isocurvature mode is modest. \textbf{Right:} The cosmic shear angular power spectrum. As more power comes from small scales
and low redshifts, cosmic shear is very sensitive to the additional white-noise component. Non-linear corrections are already important at
small $\ell$. For both plots, the case of vanishing isocurvature perturbations is shown in red, and blue lines indicate a $f_\mathrm{iso}=0.3$
model. Linear spectra are dashed, solid lines indicate results including non-linear corrections. In grey, we show the noise level of the experiments.}
        \label{fig:lensing}
    \end{figure}
\end{center}

%
%

\subsection{CMB lensing}
\label{sec:CMBlens}
As for cosmic shear, bundles of CMB photons are deformed by the LSS \citep[e.g.][]{hirata_reconstruction_2003,lewis_weak_2006}. The lensing signal of the
CMB is a complementary probe to cosmic shear and galaxy clustering since its weight function peaks at a higher redshift than the one of the EUCLID survey.
Assuming that the unlensed CMB is homogeneous, the lensing signal can be reconstructed. An unbiased, minimal variance estimator was constructed by \citep{hu_mass_2002,okamoto_cosmic_2003}
and we may write
\begin{equation}
\hat{C}_{\psi\psi}(\ell) = {C}_{\psi\psi}(\ell) + N_{\psi\psi}(\ell)\; .
\end{equation}
Let $\chi_*$ be the comoving distance to the last scattering surface. The lensing signal of the CMB is given by 
\begin{equation}
\psi = 2\int_0^{\chi_H} \mathrm{d}\chi W_\Psi\Phi, 
\end{equation}
where the CMB lensing efficiency function takes the form
\begin{equation}\label{eq:CMB_lens_weight}
W_\Psi(\chi) = \frac{\chi_* - \chi}{\chi_*\chi}\frac{H(\chi)}{ca}. 
\end{equation}
The angular power spectrum of the CMB lensing signal has the same structure as \cref{eq:cosmic_shear_power},
with the weight function replaced by \cref{eq:CMB_lens_weight}.


\subsection{Future constraints on isocurvature amplitude and ALP masses}
\label{sec:lss-sens}
In \cref{fig:lensing}, we show angular power spectra for an isocurvature scenario compared to a purely adiabatic one. The color code is the same as in \cref{fig:pk}.
Non-linearities have a rather small impact on the measurable fluctuations of the CMB lensing field. In contrast, the power spectrum for cosmic shear measurements is
affected very strongly by non-linear corrections. This is due to the fact that $(i)$ the noise level for cosmic shear is lower than for CMB lensing, thus accessing
smaller scales, and that $(ii)$ CMB lensing is most efficient at higher redshifts compared to cosmic shear since the source is farther away from us. 
The cyan lines show reconstruction and shape noise, respectively. Note that the reconstruction noise is estimated with the minimal variance estimator using all
non-vanishing temperature and polarization auto- and cross-spectra.

The relative change in the non-linear case is most pronounced even before the isocurvature component starts dominating the signal. For the linear prediction, we see
the characteristic flattening of the spectrum due to the white-noise power spectrum \citep{Enander_2017}. 
It is clear from \cref{fig:lensing} that the strongest sensitivity lies in the deep non-linear regime, which, as described earlier, is still not known precisely. 
In \citep{Feix2019}, HI intensity mapping was used to forecast constraints. This is very effective since the signal comes from higher redshifts where the power
spectrum is still linear on most of the relevant scales. 
A similar argument holds for CMB lensing. While non-linearities generate roughly five times as much lensing for a cosmic shear survey, the impact on the CMB lensing
signal is only about 5 per cent.

To put conservative constraints on the isocuvature component, we remove all scales influenced by non-linearities from the survey. In particular, we introduce a maximum
multipole $\ell^\mathrm{max}_i$ for each tomographic bin such that the total signal contains at most a one per cent contribution from non-linearities. 
Here, non-linearities can be defined via the non-linear scale $k_\mathrm{nl}$ where the variance of the smoothed density contrast becomes unity. In practice, we choose
a fixed scale cut $k_\mathrm{cut}$.
This ensures that, if $k_\mathrm{cut} < k_\mathrm{nl}$, we can rely on linear theory for the presented constraints. It also allows us to assume a Gaussian likelihood
for the data,
\begin{equation}
\label{eq:gauss_likelihood}
p(\boldsymbol{d}_\ell |\boldsymbol{\theta})= \left((2\pi)^n\mathrm{det}\boldsymbol{C}_\ell\right)^{-1/2}\exp\left[-\frac{1}{2}\left(\boldsymbol{d}_\ell-\boldsymbol{\mu}(\boldsymbol{\theta})\right)\boldsymbol{C}^{-1}_\ell\left(\boldsymbol{d}_\ell-\boldsymbol{\mu}(\boldsymbol{\theta})\right)\right] \;,
\end{equation}
where the data vector $\boldsymbol{d}_\ell$ consists of the power spectra estimators $\hat{C}^\alpha_{\ell}$ at each multipole with Gaussian covariance and components $\left(\boldsymbol{C}_\ell\right)_{\alpha\beta} \equiv \langle \hat{C}^\alpha_{\ell} \hat{C}^\beta_{\ell}\rangle$. Here $\alpha$ is a double index labeling the considered
probe. In practice, we sum over the entire multipole range and set the noise in the $i$-th bin to infinity if $\ell >\ell^\mathrm{max}_i$.
Constraints assume a sky fraction of $f_\mathrm{sky} = 1/3$ for the LSS surveys and $f_\mathrm{sky} = 0.8$ for the CMB survey. The galaxy sample is split up into 6 tomographic
bins, with an equal amount of galaxies in each bin.

\begin{center}
\begin{figure}
\centering
\includegraphics[width = 0.55\textwidth]{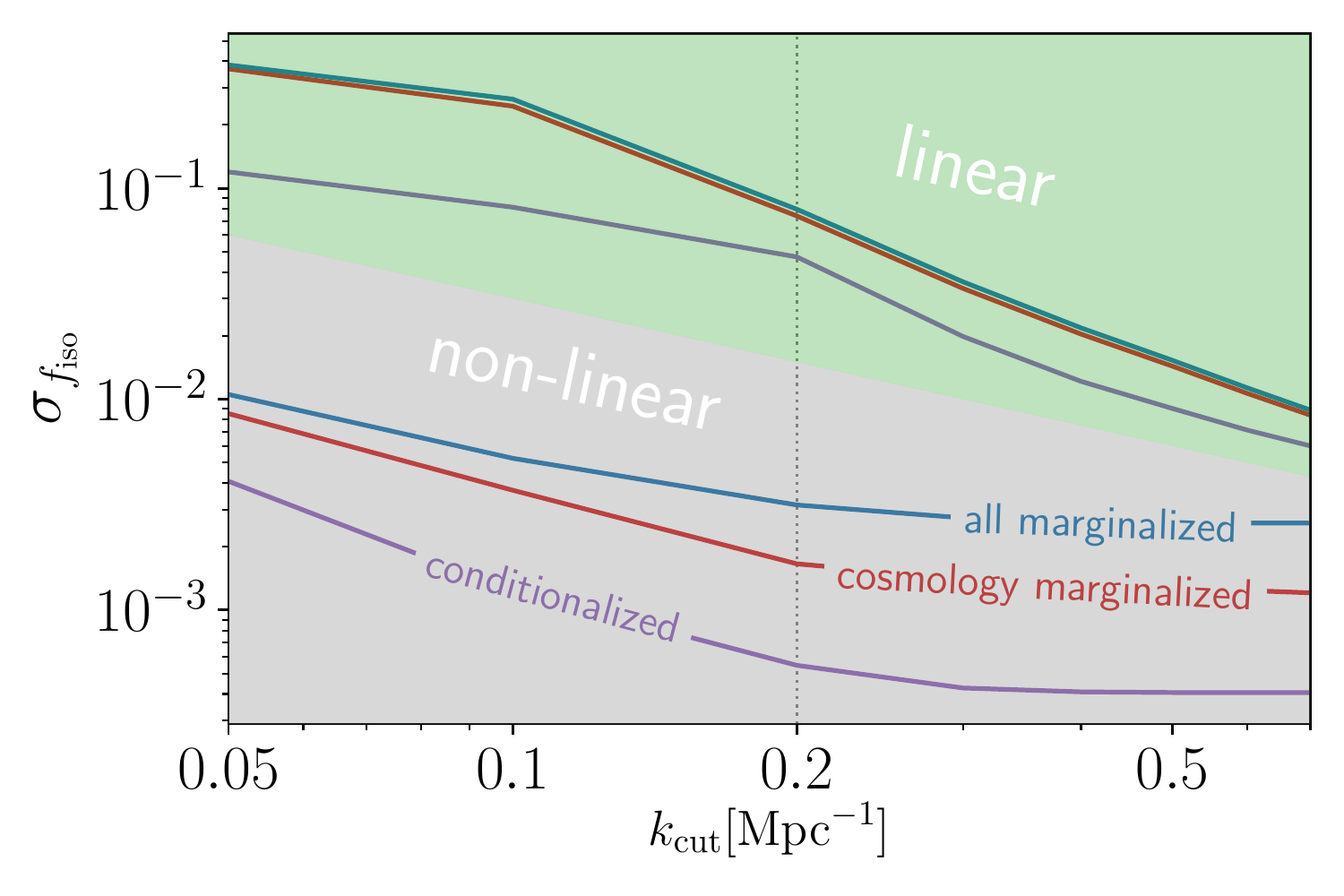}
\caption{1$\sigma$-uncertainty on $f_{\rm iso}$ as a function of the cut-off wavenumber $k_\mathrm{cut}$ obtained from combining cosmic shear,
galaxy clustering and CMB lensing. The purple line (``conditionalized'') corresponds to a setting where all cosmological and nuisance parameters
(e.g., galaxy bias) are known and fixed to their fiducial values. In red, we marginalize over cosmological parameters, but keep the bias parameters
fixed. The blue line shows the fully marginalized constraints over both cosmology and bias parameters. The conservative limits in the upper half of
the plot correspond to predictions from the linear power spectrum and the more stringent constraints in the lower part come from non-linear
predictions with higher signal-to-noise. The optimistic forecast (\textit{benchmark II}; discussed in the text) uses non-linear predictions up to a
scale of $k_\mathrm{max} = 0.2 ~ \mathrm{Mpc}^{-1}$ whereas the other forecasts use only scales up to $k_\mathrm{max} = 0.05 ~ \mathrm{Mpc}^{-1}$.}
        \label{fig:constraints_2cuts}
    \end{figure}
\end{center}

\Cref{eq:gauss_likelihood} is turned into a posterior by virtue of Bayes' theorem and approximated as a Gaussian posterior. The Fisher matrix can be constructed from the
spectra only as described in \cite{Feix2019}.
Throughout, we will assume a flat prior for all parameters. However, there are a couple of caveats in place: $(i)$ The Cramer-Rao bound does not necessarily hold for
constrained parameter spaces since it requires that the joint distribution of parameters and data and its derivatives exists for all combinations of data and parameters.
This can already be understood from the fact that the posterior will be necessarily non-Gaussian if, for example, the Fisher matrix is evaluated close to $f_\mathrm{iso} = 0$
since $f_\mathrm{iso} \geq 0$. $(ii)$ If the experiment is not very constraining, a flat prior might not be the least informative choice. In fact, the Jeffrey prior, defined in
\cref{eq:Jeffrey_prior} is the least informative prior, provided asymptotic normality has been reached. The Jeffrey prior is thus given by the Fisher information and its
dependence on the model parameters \citep{schafer_describing_2016}. \Cref{eq:Jeffrey_prior} holds for a single parameter. For higher dimensions, the prior can be constructed
sequentially. 

\begin{center}
\begin{figure}
\includegraphics[width = 0.48\textwidth]{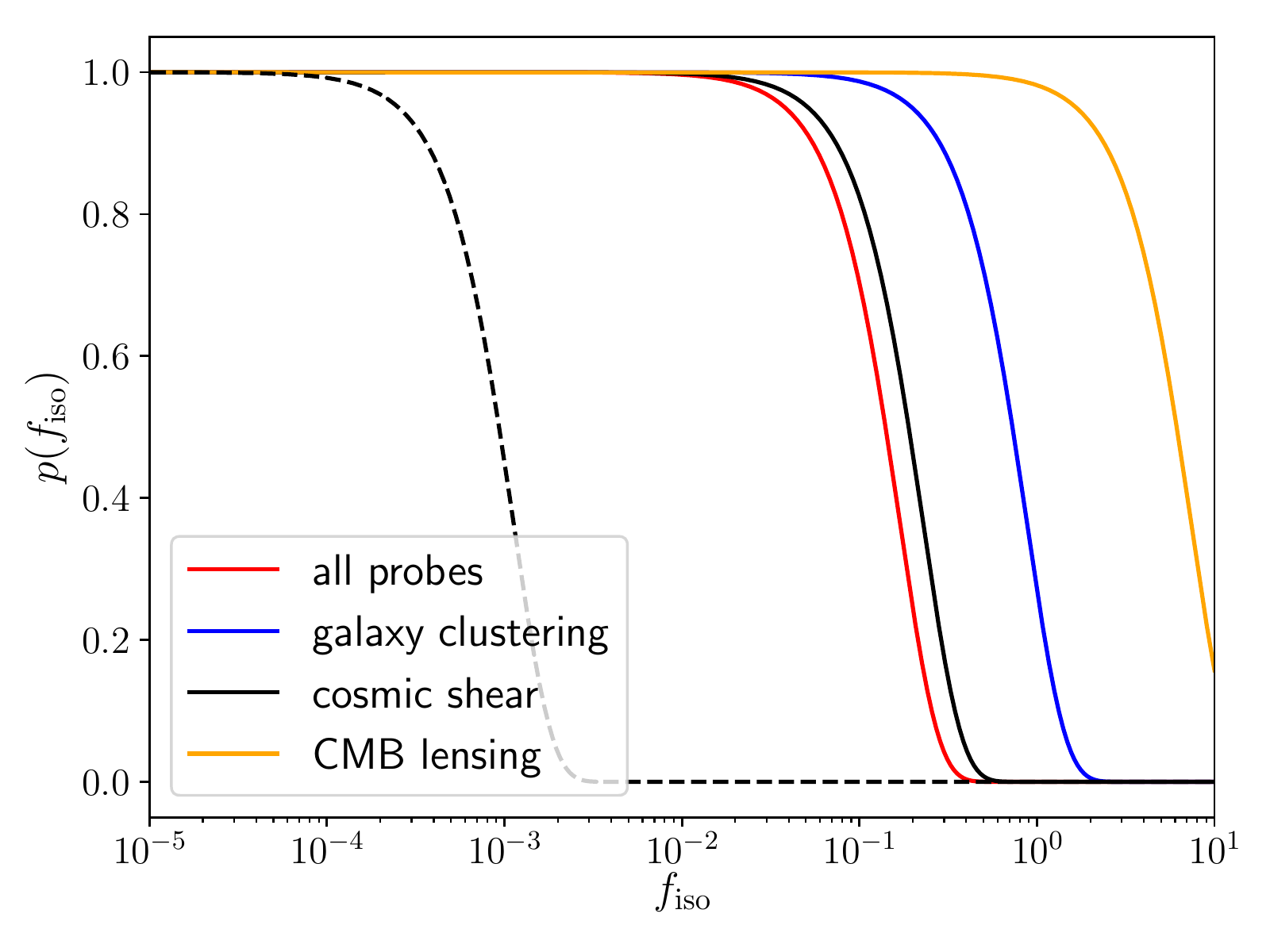}
\hfill
\includegraphics[width = 0.48\textwidth]{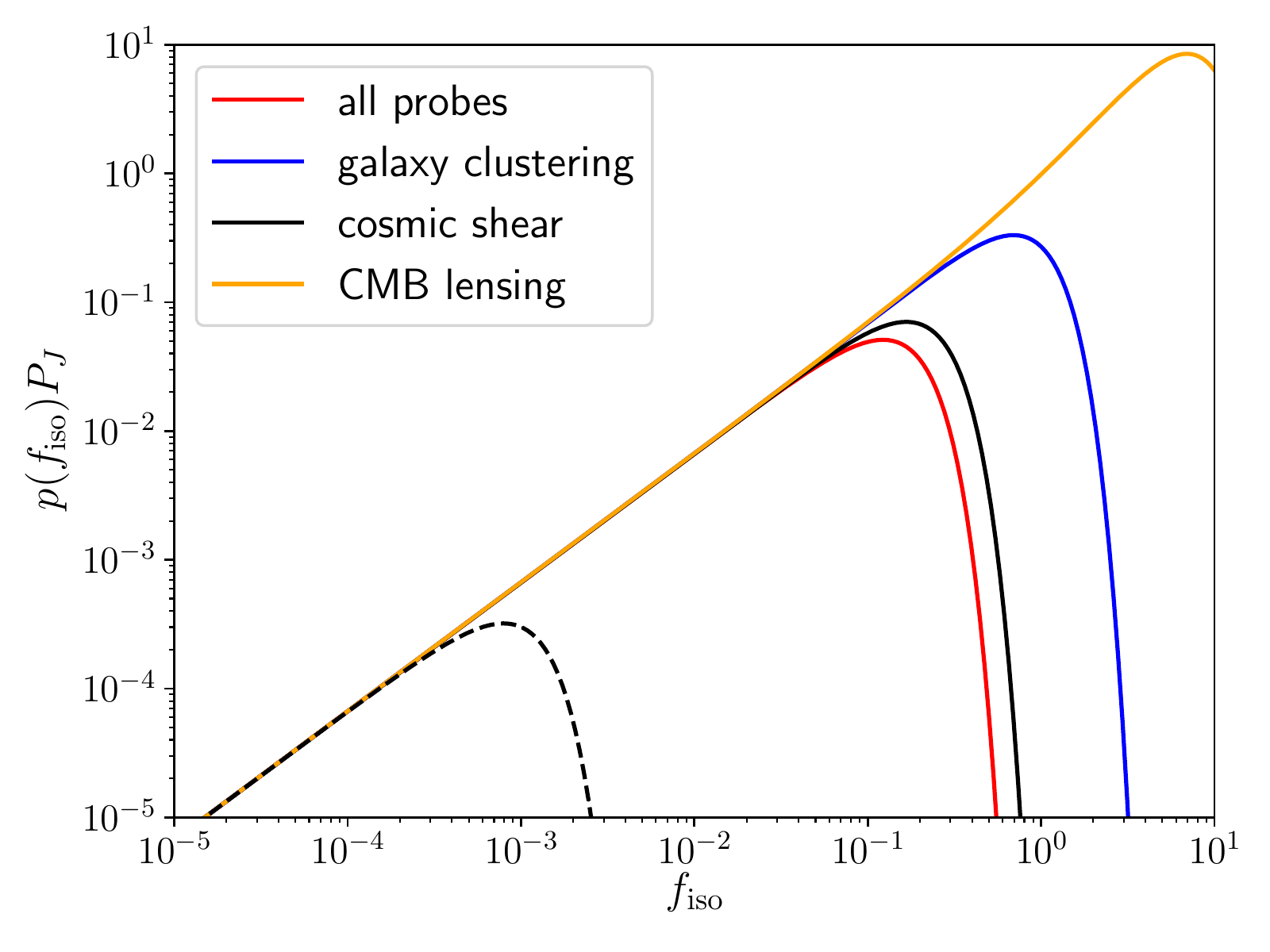}
\caption{Marginal 1$\sigma$-constraints for all probes combined (red), galaxy clustering (blue), cosmic shear(black) and CMB lensing (yellow)
in the conservative scenario (see \cref{fig:constraints_2cuts}). The dashed black line corresponds to a very optimistic situation with a correct
model for non-linearities using all probes and scales. In the left plot, we assume a flat prior whereas the Jeffrey prior is assumed for
$f_\mathrm{iso}$ in the right graph.}
        \label{fig:1dconstraints}
    \end{figure}
\end{center}

In \cref{fig:constraints_2cuts}, we present 1$\sigma$-constraints on the isocurvature component as a function of the cut-off scale. Furthermore, we show three lines with
different degrees of marginalization over cosmological and nuisance parameters to illustrate their impact. We also show the constraints obtained from using the linear (dashed)
and non-linear (solid) power spectrum.
We clearly see that the marginalization over the cosmological parameters strongly reduces the constraints on $f_\mathrm{iso}$.
The reason is that the additional isocurvature component is, to some degree, degenerate with cosmological parameters as already discussed in \cref{sec:constraints}.
In contrast, the marginalization over the bias parameters is not affected by this problem and does not change the overall constraints too much. However, this may change
when a perturbative bias expansion for the non-linear galaxy power spectrum is considered as the combination of different bias terms can give rise to a signal similar
to the isocurvature component. A further uncertainty is the shot-noise component which can be non-Poissonian. Since a non-vanishing isocurvature component in our scenario
exactly mimics a shot-noise term, there will be a strong degeneracy.
Moreover, non-linear corrections have already a strong impact at low wavenumbers (see \cref{fig:pk}). For the non-linear power spectrum, we see a flattening of the
constraints at high wavenumbers where the signal is to weak to outweigh the shot-noise component. Finally, we highlight three benchmark scenarios: \textit{conservative},
\textit{benchmark I}, and \textit{benchmark II}, where the latter uses a scale cut $k_\mathrm{cut} = 0.2\;\mathrm{Mpc}^{-1}$ and the other two use $k_\mathrm{cut} = 0.05\;\mathrm{Mpc}^{-1}$.
The difference between the \textit{benchmark I} and the \textit{conservative} scenario is that the latter uses the linear power spectrum only, which we know to be a quite
accurate on scales $k < 0.05\;\mathrm{Mpc}^{-1}$. For additional details, we refer to the discussion in \cref{sec:alp_cdm_approx}.
Comparing this to current constraints, we find that a conservative scenario with $k_\mathrm{cut}=0.05\;\mathrm{Mpc}^{-1}$ is competitive with CMB measurements (also, see
\citep{Feix2019}). Accounting for non-linear corrections could potentially improve the measurement by an order of magnitude or more.

\Cref{fig:1dconstraints} shows the marginalized 1D constraints. In the left panel, the posterior for the different probes with a flat prior is shown. For the solid lines,
the cut-off is set to $k_\mathrm{cut}=0.05\; \mathrm{Mpc}^{-1}$, corresponding to the \textit{conservative} scenario. The black dashed line shows a very optimistic setting
where we assume to have a model for non-linearities and include all scales up to $\ell = 3000$. This improves the constraints by more than two orders of magnitude. It is
obvious from the plot that cosmic shear puts the most stringent constraint on $f_\mathrm{iso}$. However, it should e noted that a more rigorous treatment of intrinsic alignments using non-linear models can change the results for cosmic shear. In our analysis we only work with two alignment parameters, describing the coupling strength of the galaxy shapes with the tidal field and the tidal torque \cite{tugendhat_angular_2017,tugendhat_statistical_2018}.
In the right panel, we show the same constraints using the least informative prior on $f_\mathrm{iso}$. As discussed before, \cref{eq:Jeffrey_prior} allows for an objective
definition of such a prior. In particular, we find that the Fisher information depends linearly on $f_\mathrm{iso}$ for $f_\mathrm{iso}\ll 1$. This reflects the fact that
no experiment will be able to detect an isocurvature mode with an amplitude very close to zero, resulting in an upper limit on $f_\mathrm{iso}$ and thus a lower limit on $m_a$.
The second effect is a shift of the bound to slightly higher values of $f_\mathrm{iso}$, with larger impact for less constraining experiments. For instance, the right wing
of the dashed black line is very close to be unaffected by the prior, as can be seen by comparing the two figures.

\begin{center}
\begin{figure}
\includegraphics[width=0.44\textwidth]{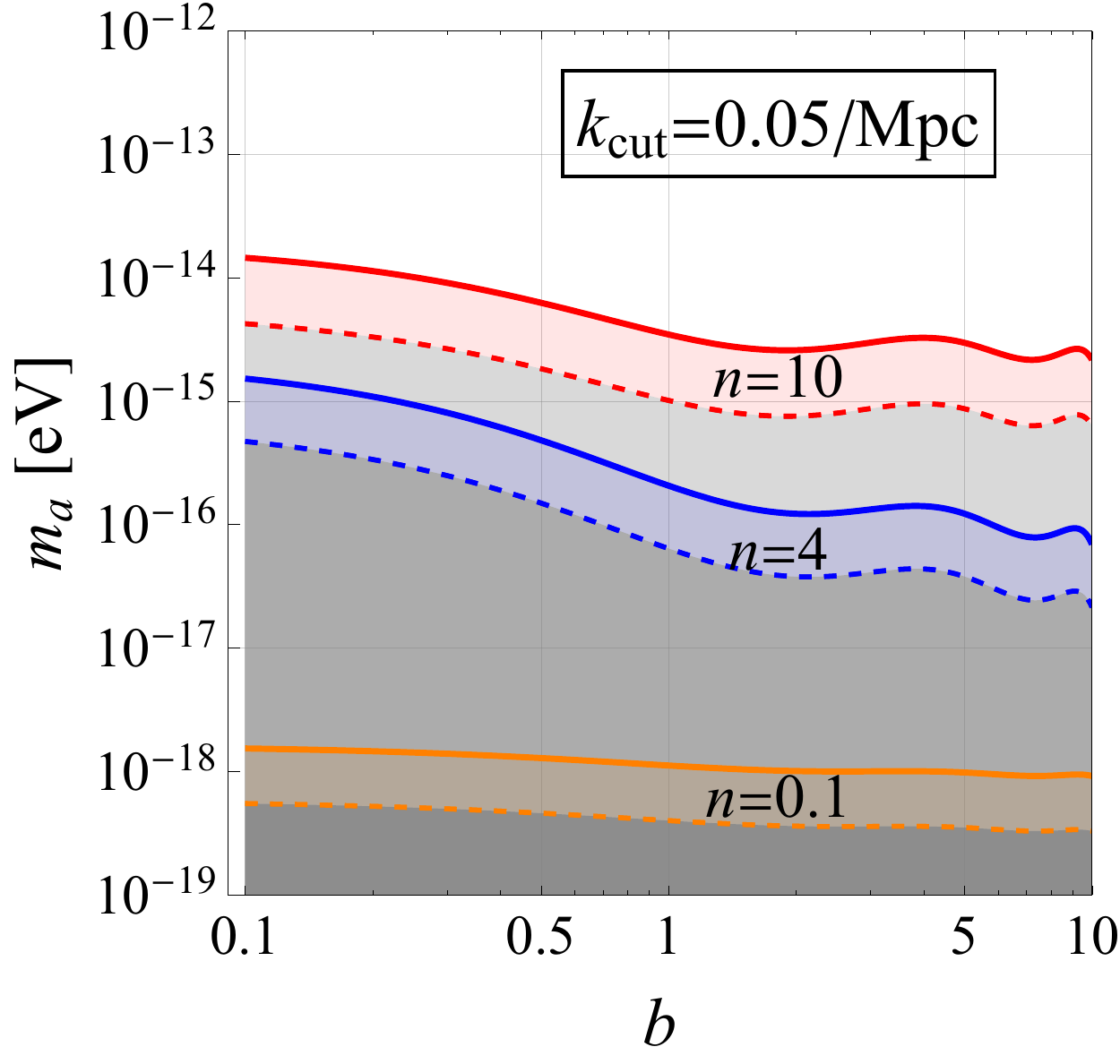}
\hfill
\includegraphics[width=0.44\textwidth]{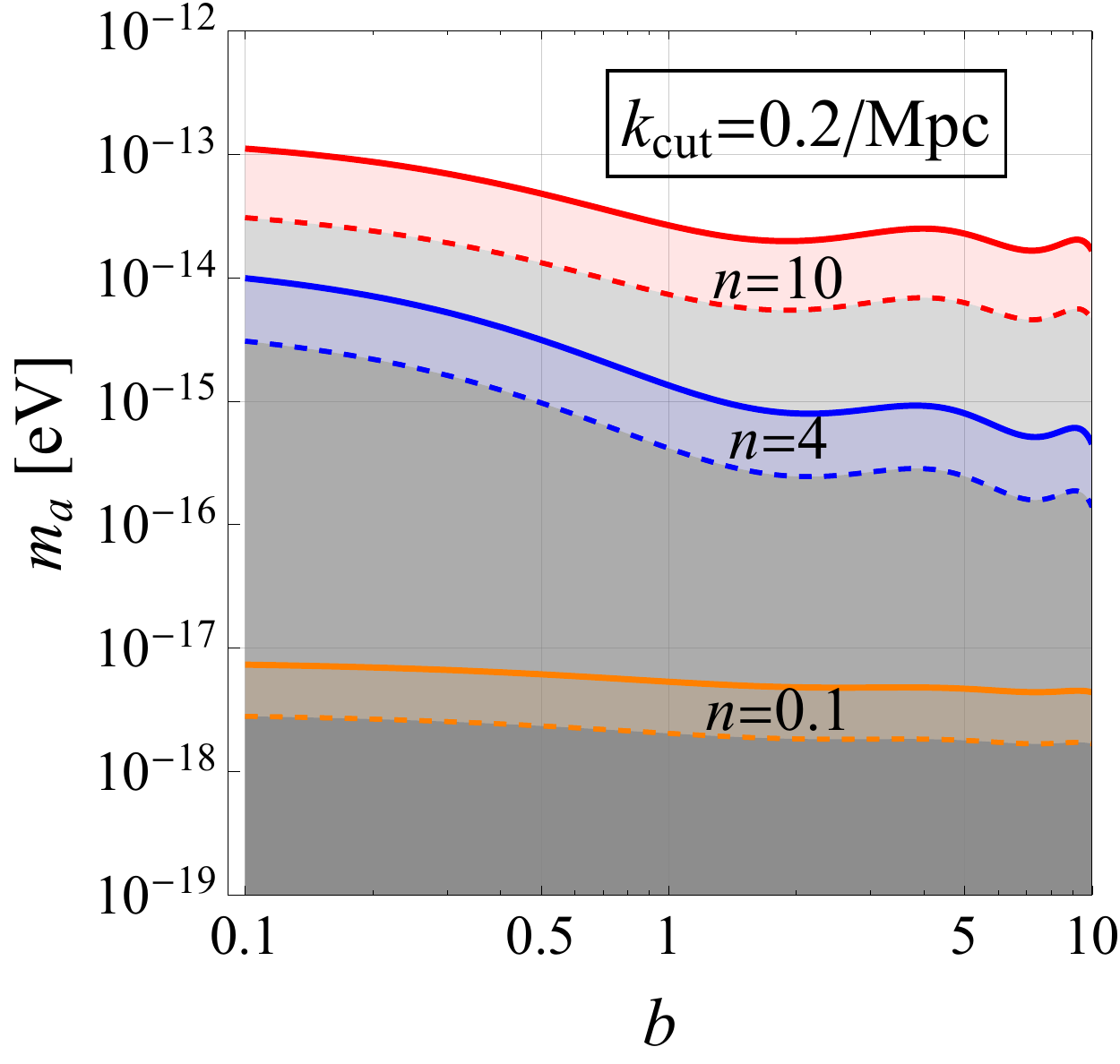}
\caption{Combined constraints (cosmic shear + clustering + CMB lensing) on the zero-temperature ALP mass, $m_{a}$, for different
assumptions on the mass-temperature dependence as given in \cref{Eq:MassParameterization}. Solid curves show estimated bounds and
bands between solid and dashed curves indicate a factor 5 uncertainty.
The case $n=0.1$ is numerically very similar to the case
    of a temperature-independent ALP mass with $n=0$.
The left panel corresponds to \textit{benchmark I} with $k_{\rm cut} = 0.05$ Mpc$^{-1}$
and	$f_{\rm iso} < 0.011$ ($1\sigma$-level). More optimistic results are shown in the right panel for the \textit{benchmark II} scenario with $k_{\rm cut} = 0.2$ Mpc$^{-1}$
and $f_{\rm iso} < 0.003$ ($1\sigma$-level). The region below the curves is disfavored and ALPs are assumed to provide all DM.}
\label{fig:mass_bounds}
\end{figure}
\end{center}

Finally, in \cref{fig:mass_bounds}, we express the constraints on
$f_\mathrm{iso}$ in terms of the axion mass $m_a$ and the two other
parameters, $n$ and $b$, controlling the mass-temperature dependence.
Assuming that all DM has been produced after inflation, the three
parameters can be directly linked to the relative amplitude of the
isocurvature component at the pivot scale. The two plots show
constraints for different assumptions concerning the scale up to which
non-linearities can be modeled reliably in $\Lambda$CDM. The conservative
case corresponds to \textit{benchmark I} (left panel).
This leads to a bound $f_{\rm iso} < 0.011$ at the $1\sigma$-level,
and the resulting lower bounds on the ALP mass are
competitive with constraints from CMB stage-IV and intensity mapping
experiments presented in \cite{Feix2019}.
Corresponding to \textit{benchmark II}, the right panel shows more optimistic
results that lead to $f_{\rm iso} < 0.003$ ($1\sigma$), and thus an improvement
by an order of magnitude for the bounds on $m_a$. Given the current limit from
\cref{eq:bounds} and the constraints from current observations shown in
\cref{fig:mass_bound_current}, we see that potentially very large
regions of parameter space can be tested with those observations.
The expected bounds are competitive with the limits derived in
\citep{Irsic2019} from reionization and Lyman-$\alpha$ observations.

\section{Conclusions}
In this paper, we studied the sensitivity of CMB and LSS experiments on specific isocurvature fluctuations that can originate from ALPs
produced after inflation. 
The isocurvature component manifests as an increase in power on smaller scales due to its white-noise character. If all
DM exists in the form of ALPs, upper bounds on the amplitude of the isocurvature component can be translated into lower bounds on the ALP mass, with
some model dependence on the exact emergence of the ALP mass as a function of temperature. For details on the ALP model assumptions, we refer to \cite{Feix2019}.
We used current primary anisotropies of the CMB \cite{Planck_likelihood_2019}, the abundance of galaxy clusters detected through their SZ signal by
Planck \cite{PlanckSZ_15} and BAO measurements from 6dFGS \cite{Beutler:2011hx}, SDSS-MGS \cite{Ross:2014qpa}, and BOSS DR12 \citep{Alam:2016hwk} to
obtain constraints on the isocurvature component and ALP mass. Lastly, we investigated possible constraints when combining a CMB stage-IV experiment
with a Euclid-like survey, using cosmic shear and galaxy clustering. 
We summarize our main results as follows:
\begin{enumerate}[i)]

\item There exists a preference for a non-vanishing white-noise
  isocurvature component, characterized by the parameter $f_\mathrm{iso}$
  in \cref{eq:fiso}, in the temperature and polarization data of the
  CMB (3.1$\sigma$).

\item When allowing for an arbitrary isocurvature tilt, the CMB data
  prefers the white noise spectrum predicted by the post-inflationary ALP
  scenario. In this case, however, the amplitude is consistent with
  zero within 2$\sigma$.

\item While the ALP model explored here fits a real feature present in
  the primary CMB temperature and polarization data (see above), we
  caution not to over-interpret this as the signature of an isocurvature
  mode. At the moment, this is discouraged by LSS data such as galaxy
  clusters. Galaxy clusters combined with BAO alone set an upper bound
  of $f_{\rm iso} < 0.27 \, (0.53)$ at the 1$\sigma$-level (95\%~CL).

\item Combining CMB, galaxy clusters and BAO, we find $f_{\rm iso} <
  0.64$ at 95\%~CL, which can be translated into a lower bound on the
  ALP mass ranging from $m_a \gtrsim 10^{-21}$~eV for a weak ALP
  mass-temperature dependence up to $m_a \gtrsim 10^{-17}$~eV for a
  strong mass-temperature dependence.

\item The sensitivity of future LSS experiments strongly depends on
  the modeling of non-linear structure formation. When removing most
  of the non-linearities from the survey, a Euclid-like setting will
  be able to constrain $m_a\gtrsim 10^{-18}$ to $10^{-13}$~eV
  (depending on the mass-temperature dependence), which is competitive
  with CMB stage-IV and HI intensity mapping experiments \cite{Feix2019}.
  When mildly non-linear scales are included, the bound improves by an
  order of magnitude.

\end{enumerate}

\label{sec:conclusion}
\acknowledgments 
SH would like to thank Martina Gerbino for very helpful discussions about the Planck likelihood code.
This research was supported by the Excellence
Initiative of the German Federal and State Governments at Heidelberg
University, by the European Union's Horizon 2020 research and
innovation programme under the Marie Sklodowska-Curie grant agreement
No 674896 (Elusi$\nu$es), and by the Heidelberg Karlsruhe Research
Partnership (HEiKA). SH acknowledges support from the Vetenskapsr\r{a}det (Swedish Research Council) through contract No. 638-2013-8993 and the Oskar Klein Centre for Cosmoparticle Physics.

\bibliographystyle{JHEP}
\bibliography{axion_cosmo}

\end{document}